\documentclass[12pt]{article}

\usepackage{amsmath}
\usepackage{slashed}
\usepackage{amsfonts}
\usepackage{amssymb}
\usepackage{graphicx}
\usepackage{grffile,cite}
\usepackage{subfig}
\usepackage{hyperref}
\usepackage{mathtools}
\usepackage{tikz}
\usepackage{xcolor}

\usepackage{float}

\begin{document}
\thispagestyle{empty}

\begin{center}

%\vspace*{0.5cm}
{\LARGE\bf{Crossing symmetry including non planar diagrams in perturbative QFT}}
\bigskip

{\large Ritabrata Bhattacharya}\, \, \\

\bigskip 
\bigskip

{\small
\textit{Indian Institute of Science Education and Research Bhopal},
IISERB, Bhauri, Bhopal - 462066, Madhya Pradesh, India\\[3mm]
}

\end{center}
\begin{center}
\it{ritabratabhattacharya@iiserb.ac.in}
\end{center}
\bigskip 
 %
%
%\vspace*{1cm}
\begin{center} 
{\bf Abstract} 
\end{center}
%\vspace*{-0.35in}

\begin{quotation}
\noindent 
We venture a proof of crossing symmetry for non-planar diagrams in perturbative QFT. For the planar diagrams a proof of crossing is available in the literature and our method closely follows the one depicted in that case. We classify the non-planar diagrams broadly into two types. For one of these types the proof is pretty straightforward and hence the result extends to all point all loop on-shell amplitudes. These are called the ``trivial" cases while for the other type we find certain cases called the ``non trivial" cases for which the proof is much more subtle. We present an explicit example of such a ``non trivial" case at 3-loop order and argue how the proof of crossing symmetry holds true when all subtleties are taken into consideration. Based on this simple example we argue how the proof works out in general for these ``non-trivial" cases at higher loop and with arbitrary number of non-planar edges.
{\small

}
\end{quotation}

\newpage
\tableofcontents
\setcounter{footnote}{0}

\section{Introduction}
From the known results of Quantum Field Theory (QFT) we observe that by different scattering experiments one can never distinguish between particles and anti-particles with the opposite energy and momentum \cite{Stueckelberg:1941}. Theoretically this fact translates to the statement of \textit{crossing symmetry} which says that on-shell scattering amplitudes for processes involving the particle and the anti-particle are boundary or limiting values of one and the same analytic function, regardless of the number and type of the remaining particles it interacts with. This fundamentally Lorentzian notion is believed to be a reflection of compatibility of quantum theory with causality, locality and unitarity.

At its heart crossing symmetry is an analytic property of the S-matrix. Since its introduction in 1954 by Gell-Mann, Goldberger and Thirring \cite{Gell-Mann:1954}, establishing crossing symmetry as a consequence of the aforementioned physical principles has remained an open problem. Meanwhile the works of Bros, Epstein and Glaser \cite{BEG:1964, BEG:1965, BEG:1972, Bros:1986} shed some light on a way to prove crossing symmetry in the framework of axiomatic quantum field theory in the Lehmann-Symanzik-Zimmerman (LSZ) formalism \cite{LSZ:1955}. This technique applies the apparatus of complex analysis in multiple variables overcoming the absence of a convincing physical explanation. Due to the LSZ approach their study involves working with off-shell Green's functions and inferring properties of scattering amplitudes by extensive use of analytic completion theorems in the on-shell limit. Similar results in the context of off-shell Green's function in String Field Theory (SFT) follow from the seminal work of de Lacroix, Erbin and Sen \cite{DeLacroix:2018arq} (see also \cite{Bhattacharya:2020gar} for a partial generalisation). At this point let us discuss the advantages and limitations of this approach.
\begin{itemize}
\item{\textit{Advantages}:\\
This approach uses only some basic axioms which is why there results are valid for both perturbative as well a non-perturbative QFTs. Also the analyticity properties that are used for their results are mathematically quite robust and can be used to prove other analytic properties of the Green's functions such as spectral decomposition etc. as well.}
\item{\textit{Limitations}:\\
The process of analytic continuation in this treatment is rather abstract and difficult to interpret in terms of particle scattering. This fact along with the use of off-shell quantities creates an obstruction to understanding the physical origin of crossing symmetry and the process itself turns out to be technically strenuous. As a consequence, proofs of this type have been completed only in theories without massless particles in the case of 2 $\rightarrow$ 2 \cite{BEG:1964, BEG:1965} and 2 $\rightarrow$ 3 scattering \cite{BEG:1972, Bros:1986}. For reviews, see \cite{Epstein:1966, Sommer:1970, BLOT:1989, Bros:1980}.}
\end{itemize}
In contrast the approach to proving crossing symmetry adopted in \cite{Mizera:2021fap} is much more straightforward and the process of analytic continuation is done, keeping the on-shell condition as well as the momentum conservation intact throughout the process. This makes the understanding of the origin of crossing symmetry much clear. Although, the method itself is limited to the case of perturbative QFT only, it does have the added advantage that it generalises easily to the case of arbitrary $n$-point functions (with $n\geq4$). In \cite{Mizera:2021fap} the proof of crossing symmetry was limited to the case of planar diagrams.

In this article we attempt to go beyond the planar limit and use the technique of \cite{Mizera:2021fap} in the context of non-planar diagrams as well. It is best at this stage to inform the reader about the scope of this paper. We explicitly work out different 4-point non-planar amplitudes at 2-loop and 3-loop order and this gives us the essential insights that are necessary. These results can be generalised to the case of arbitrary $n$-point functions (with $n\geq4$). The different non-planar diagrams are broadly separated into two types namely,
\begin{enumerate}
\item{\textit{Diagrams with number of non-planar internal edges = 1}}
\item{\textit{Diagrams with number of non-planar internal edges $>$ 1}}
\end{enumerate} 
For the first type of diagrams we show that they respect crossing symmetry for arbitrary number of external states and to all loop order. For the second type we identify the cases where we can potentially run into a singularity during analytic continuation. In this context we work concretely with the specific example of a 4-point function at 3-loop order with 2 non-planar edges. We argue that the singularity it can run into is a sub-leading Landau singularity i.e. there is no leading Landau singularity encountered during the analytic continuation. Since we are concerned only with the leading Landau singularities this implies that the diagram is crossing symmetric. Finally, based on this proof we provide a general argument for higher loop order with arbitrary number of non-planar edges as to how we can carry out the analytic continuation from one channel to its crossed channel without any obstruction (at least for $D>2$).

The rest of the paper is organised as follows. In $\S$\ref{sec:Review} we briefly review the notations and results of \cite{Mizera:2021fap}. In particular we talk about the analytic continuation via five steps keeping the on-shell condition along with the overall momentum conservation fixed. We also present the expression of the internal momenta in terms of the Schwinger parameters after integrating out the loop momenta variables. In $\S$\ref{sec:NP_main} we present our main results of the paper. We present the explicit proof of crossing symmetry for non-planar diagrams at 2-loop order. We then work at 3-loop order leaving some explicit examples of the ``trivial" cases for appendix \ref{app:A}. In the main body of $\S$\ref{sec:NP_main} we present in full detail a specific example of the ``non trivial" case that requires a more careful attention and then give a general argument as to how the proof works out for arbitrary loop order with arbitrary number of non-planar edges. We also mention the way to generalise our results from the case of 4-point functions to higher point by the use of the method presented in $\S$\ref{sec:General}. Finally we conclude the article with $\S$\ref{sec:Conclusion} summarising the results of the paper and some future directions.

\section{Review of crossing in planar limit}
\label{sec:Review}
\begin{figure}[h!]
\centering
\includegraphics[scale=0.5]{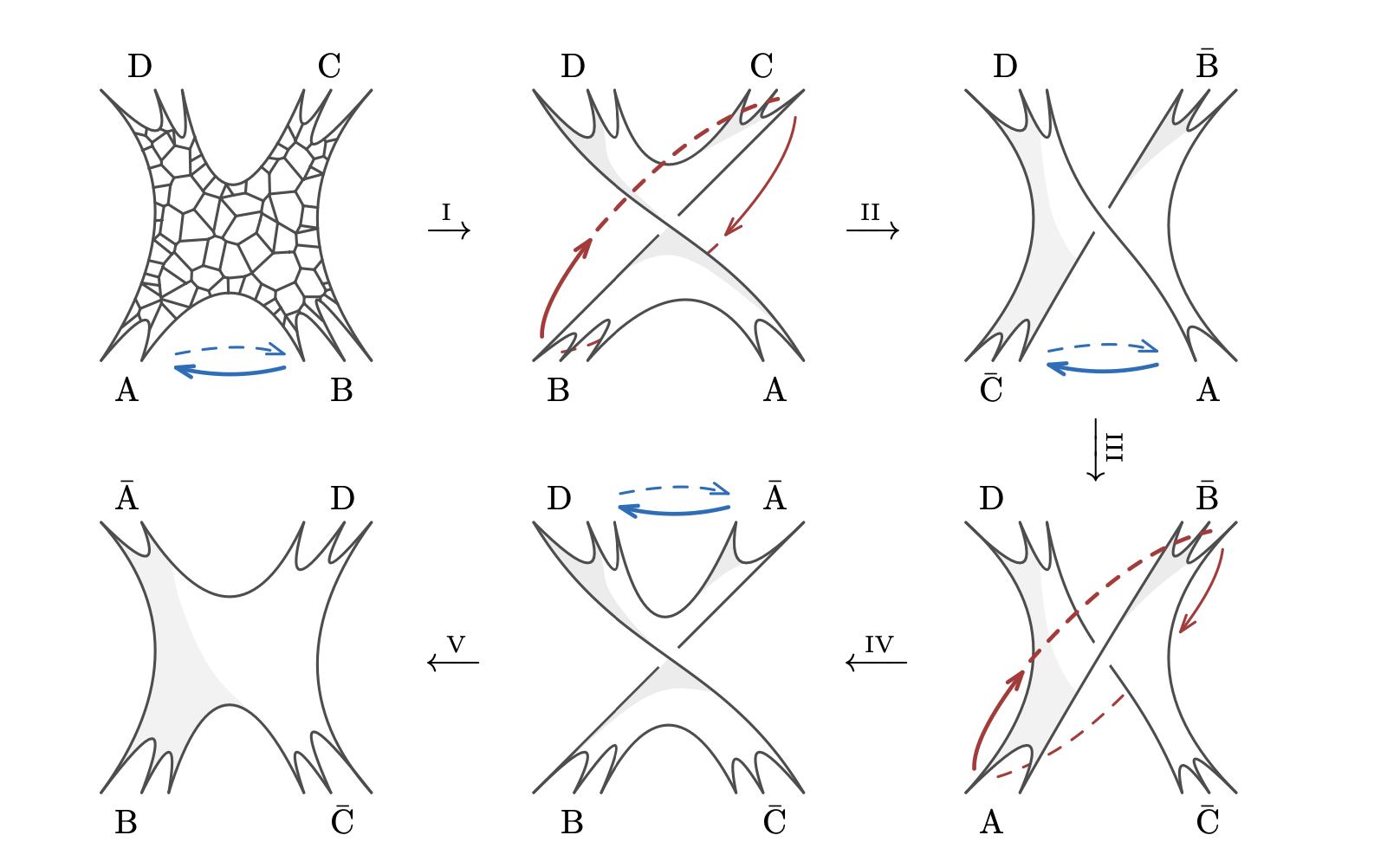}
\caption{Five steps to carry out the analytic continuation from one configuration to its crossed configuration i.e. from $AB\rightarrow CD$ to $B\bar{C}\rightarrow D\bar{A}$. (We acknowledge this figure to be an exact replica of figure 1 of \cite{Mizera:2021fap}.)}
\label{fig:1}
\end{figure}

Here we briefly review the work \cite{Mizera:2021fap} as the results of this paper closely follows the concepts and some of the results stated there. The proof of crossing in the planar limit for arbitrary number of external particles and loops comprises of five steps each of which involves either the deformation of loop integration contours or the deformation/analytic continuation of the external momenta preserving the on-shell condition and the overall momentum conservation. The idea is that one can carry out this five steps to go from one configuration to the crossed configuration by avoiding all the singularities of the loop integrand. This establishes the fact that the result of the scattering amplitude in two crossed channels are obtained by different limits of the same analytic function, which is exactly the statement of crossing symmetry. 
\begin{equation}
\mathcal{T}_{AB\rightarrow CD}\equiv \mathcal{T}_{B\bar{C}\rightarrow D\bar{A}}
\end{equation}
where the $\equiv$ denotes that both sides are the results at different limits of the same analytic function. 
The five steps mentioned above are depicted pictorially in Figure \ref{fig:1} which is reproduced exactly as in \cite{Mizera:2021fap}. 
All one needs to establish is that during each of this step we can deform the loop integration contour or the external momenta in such a way that we can always avoid the singularities of the integrand. In the rest of the section we will first discuss the locations and the physical meaning of these singularities by the use of Landau equations and then describe how to avoid these singularities when carrying out the five steps discussed above.

\subsection{Landau equations} 
We will be working in the framework of perturbation theory where the scattering amplitudes can be expressed as a perturbation series over the loop diagrams. In this context the \textit{Landau equations} \cite{Landau:1959fi} provide a set of algebraic conditions which determine the positions of the singularities of the loop integrand. As we will see, the singularities have the straightforward interpretation of some intermediate particle going on-shell. 

In our convention we take all momentum to be ingoing for the Feynman diagrams so that entities with energy $E=p^0>0$ are labeled as the incoming particles and anti-particles while entities with energy $E=p^0<0$ are labelled outgoing particles and anti-particles. We work in a D dimensional spacetime with the Minkowski metric ($+,-,-,-,\dots$). Consider a scalar Feynmann diagram containing `$n$' external particles or ``legs", `$L$' number of loops and `$E$' number of internal propagators or ``edges". The result is of the following general form,
\begin{equation}
A(p_1,p_2,\dots,p_{n-1})=\int \prod_{I=1}^L d^D\ell_I\ \prod_{e=1}^E \frac{i\hslash}{(q_e^2-m_e^2+i\varepsilon)}\ .
\label{eq:amp1}
\end{equation}   
Here we have used the overall momentum conservation i.e. $p_1+p_2+\dots+p_{n-1}+p_n=0$ as a result of which the amplitude is a function of $n-1$ independent external momenta. The $\ell_I$'s denote the loop momenta whereas, $q_e$ and $m_e$ denote the total momentum and the mass of the intermediate particle flowing through the $e$-th edge respectively. The $i\varepsilon$ ensures that causality is maintained and the expression should be understood as the limit where $\varepsilon\rightarrow 0^+$. 

We can rewrite \eqref{eq:amp1} by using the so called Schwinger parametrization viz.,
\begin{equation}
\frac{i\hslash}{(q_e^2-m_e^2+i\varepsilon)}=\int_0^\infty d\alpha_e\ e^{\frac{i}{\hslash}(q_e^2-m_e^2+i\varepsilon)\alpha_e}\ ,
\label{eq:SchPar}
\end{equation} 
as the following expression,
\begin{equation}
A(p_1,p_2,\dots,p_{n-1})=\int \prod_{e=1}^E d\alpha_e\ \prod_{i=1}^L d^D\ell_I\ e^{\frac{i}{\hslash}(\mathcal{V}+i\varepsilon\sum_{e=1}^E\alpha_e)}\ .
\label{eq:amp2}
\end{equation}
Here we have 
\begin{equation}
\mathcal{V}=\sum_{e=1}^{E}\alpha_e(q_e^2-m_e^2)\ ,
\end{equation}
which is sometimes called in the literature as the \textit{world-line action}. Think of any Feynman diagram with $n$ external legs. All the internal edges always carry total internal momenta which is linear in both the external as well as the loop momenta. As a result of momentum conservation at each vertex of the diagram we have ($\forall\ \mu=0,1,\dots, D-1$),
\begin{equation}
P_{v}^{\mu}+\sum_{e=1}^{E}\eta_{ve}q_e^{\mu}=0\ ,\ \ \forall\ v.
\label{eq:LE_1}
\end{equation}  
Here $P_v$ denotes the total external momenta entering the vertex $v$ whereas $\eta_{ve}$ takes the value +1 ($-$1) if the $e$-th edge is incoming (outgoing) at the vertex $v$ and 0 otherwise. Let us now turn our attention to the expression \eqref{eq:amp2} and notice that the most obvious singularities of the integrand are associated to the saddle points in the classical limit, $\hslash\rightarrow 0$, inside the integration domain. They can be found by extremizing $\mathcal{V}$ with respect to the $\ell_I$'s and the $\alpha$'s. Thus we get,
\begin{eqnarray}
\frac{\partial\mathcal{V}}{\partial\ell^{\mu}_I}=0 &\Rightarrow& \sum_{e=1}^E\eta_{Ie}\alpha_e q^{\mu}_e=0\ ,\ \ \forall\ I=1,2,\dots,L , \label{eq:LE_2}\\
\frac{\partial\mathcal{V}}{\partial\alpha_e}=0 &\Rightarrow& q_e^2-m_e^2=0\ ,\ \ \forall\ e=1,2,\dots,E\ . \label{eq:LE_3}
\end{eqnarray} 
Note that $\eta_{Ie}$ takes the value $+1\ (-1)$ if the $I$-th loop momentum flows in the same (opposite) direction as $q_e$ and 0 otherwise. The equations \eqref{eq:LE_1}, \eqref{eq:LE_2} and \eqref{eq:LE_3} are together known as the \textit{Landau equations} of which the first two are sometimes clubbed separately as the \textit{linear Landau equations}. It is the last one i.e. equation \eqref{eq:LE_3} from which we can see that these singularities are associated to the intermediate particles going on-shell. Now, note that the case where all the intermediate particles go on-shell together is known as the ``leading Landau singularity" and this will be the one which holds the most significance in our analysis. The cases where only a proper subset of all the internal edges go on-shell are called ``sub-leading Landau singularities" and are associated to saddle points confined to the boundaries of the integration.

\subsection{Solutions of the Landau equations}
Although the Landau equations themselves have been known to physicists for more than half a century, general solutions to these equations are understood rather poorly. Multiple textbooks have been written on this subject from the perspective of combinatorics \cite{Nakanishi1971}, complex analysis \cite{Eden1966, Todorov2014}, algebraic geometry and topology \cite{HT:1966, Pham:1967, Pham:2011} and axiomatic quantum field theory \cite{Iagolnitzer:2014}. 

Firstly, note that one can write the $q_e$'s as a function of the external momenta and the $\alpha$'s when the loop momenta $\ell_I$'s are integrated over. Thus, the Landau equations turn out to be a set of algebraic equations for the $\alpha$'s. The solutions to the Landau equations can be real as well as complex. For real solutions of the $\alpha$'s, only the real positive solutions are of interest to us since from \eqref{eq:SchPar} we see that the region of integration includes only that real sub-region for which all the $\alpha$'s are positive semidefinite. These solutions are called positive-$\alpha$ or $+\alpha$-Landau surfaces and have a special meaning, because they correspond to singularities on the undistorted contour\footnote{These type of singularities are also called pinch singularities in the literature.}. These singularities have the interpretation of classical particles propagating in space time i.e. the intermediate particles going on shell with the Schwinger parameters being proportional to the proper time of the corresponding particle \cite{Coleman:1965xm}. The real positive solutions of the Landau equations will be of utmost importance to us and are reasonably well understood in comparison to complex solutions. Extensive studies on the real solutions have been done, particularly due to the work of Chandler, Iagolnitzer and Stapp from the perspective of macroscopic causality and microanalyticity \cite{Chandler:1968, Stapp:1968, Iagolnitzer:1969, IS:1969, CS:1969, BOP:1969}, cuts and discontinuities \cite{Boyling:1966, CoS:1970, CaS:1972, KS:1977}, or Steinmann relations \cite{Pham:1967_2, Boyling:1968, CaS:1975, CoS:1975}. Complex solutions are equally important for example to determine the singularities of discontinuities  
if and when they contribute. We will not discuss further about the complex solutions here (see, e.g. \cite{Andersson:1966, Halpern:1962} for partial progress).

Explicit solutions to Landau equations are limited to diagrams with small number of loops and external legs, with at least partial results up to $n\leq 5$; see, e.g. \cite{OT:1962, Islam:1966, Risk:1968}. Some more recent works (e.g. \cite{Correia:2025yao, Fevola:2023kaw, Fevola:2023fzn}) focus on developing computational tools to determine Landau singularities using different algorithms. Before ending this subsection let us note one final observation about the solutions. If we have a $+\alpha$-Landau surface at $\lbrace\alpha^*_e\rbrace$ ($e=$ 1,2,\dots, E) then there must be another solution surface at $\lbrace\lambda \alpha^*_e\rbrace$ ($e=$ 1,2,\dots, E), where $\lambda\in \mathbb{C}$. This fact implies that we cannot avoid pinch singularities simply by deforming the integration contours.

\subsection{Important results}
Let us turn our attention to \eqref{eq:amp2} and realise that since the total internal momenta $q_e$ are linear in the loop momenta $\ell_I$ the integrand takes the form of a gaussian in the $\ell_I$ variables by completing the squares appropriately. One can easily check that integrating over the loop momenta one by one we reach to the following form,
\begin{equation}
A(p_1,p_2,\dots,p_{n-1})=c\int_0^{\infty}\frac{\text{d}^E\alpha}{\mathcal{U}^{D/2}}\ e^{\frac{i}{\hslash}(\mathcal{V}+i\varepsilon\sum_{e=1}^E\alpha_e)}\ ,
\label{eq:Main_Integral}
\end{equation} 
where $\text{d}^E\alpha\equiv \prod_{e=1}^E d\alpha_e$, the constant $c$ is unimportant for the question of singularities. We have used the same symbol for the world sheet action as in \eqref{eq:amp2} since on the Gaussian saddle point they are given in terms of the same function (see, \cite{Mizera:2021fap}),
\begin{equation}
\mathcal{V}(\alpha_e)=\sum_{e=1}^E(q_e^2-m_e^2)\alpha_e\bigg|_{\eqref{eq:LE_1},\eqref{eq:LE_2}}\ .
\end{equation}
The world-line action is now a function of the Schwinger parameters and the Mandelstam invariants only. Finally, the function $\mathcal{U}$ is a homogeneous polynomial with degree $L$ in the $\alpha_e$'s,
\begin{equation}
\mathcal{U}:=\sum_{\substack{\text{spanning}\\ \text{trees }T}}\prod_{e'\notin T}\alpha_{e'}\ .
\label{eq:expressionU}
\end{equation}
By spanning trees we mean all possible connected tree diagrams with $n$ external legs, obtained by removing $L$ internal edges from the original $L$ loop diagram with $n$ external legs. On the undistorted contour we have $\mathcal{U}>0$.

At this point let us mention for the sake of completeness that for theories other than scalar theories i.e. theories with spin, only the numerator of the integrand is affected via multiplication of some polynomial in the integration variables (see \cite{Smirnov:1991} for details). As a result of this some singularities (which are basically poles of the integrand in \eqref{eq:amp1}) may be removed or the residues may change but it can never introduce new poles or singularities. So analysis of the scalar amplitude should suffice for proving crossing symmetry via analytic continuation. Also, note that for theories with spin, since the numerator is a polynomial in the loop momenta it does not change the fact that the argument of the exponential is still quadratic in the loop momenta and we get some Gaussian integral over the $\ell_I$'s. The only difference will be that we may get some higher moments of the Gaussian integral as well. 

Also for obtaining the correct causality condition that selects the correct homology class giving rise to physical amplitudes we will deform the external kinematics as well as the Schwinger parameters in such a way that,
\begin{equation}
\text{Im}\mathcal{V}>0
\end{equation}
and approach the physical regions with $\text{Im}\mathcal{V}\rightarrow 0^+$. The readers can read the section II.B.2 of \cite{Mizera:2021fap} for some more details.

Now, let us turn our attention to the general solution of $q_e$ that solves both the linear Landau equations. For any arbitrary Feynman diagram the linear Landau equations \eqref{eq:LE_1} and \eqref{eq:LE_2} can be solved explicitly and give,
\begin{equation}
q_e^{\mu}=\frac{1}{\mathcal{U}}\sum_{\substack{\text{spanning}\\ \text{trees }T}}P_{T,e}^{\mu}\prod_{e'\notin T}\alpha_{e'}\ ,
\label{eq:edge1}
\end{equation}
where $P_{T,e}^{\mu}$ denotes the total external momenta flowing through the edge $e$ along the spanning tree $T$ in the orientation of the edge. In case $T$ does not include the edge $e$ then, $P_{T,e}^{\mu}=0$. One can also check that the product of $\alpha_e$ and $q_e$ is given by,
\begin{equation}
q_e^{\mu}\alpha_e=\frac{1}{\mathcal{U}}\sum_{T=T_1\sqcup T_2\sqcup e}P_{T,e}^{\mu}\prod_{e'\notin T_1,T_2}\alpha_{e'}\ ,
\label{eq:edge2}
\end{equation}
where the sum runs over only those trees $T$ that are disjoint unions of two trees $T_1$, $T_2$ and the edge $e$ itself. It easy to check that these expressions solve \eqref{eq:LE_1} and \eqref{eq:LE_2} explicitly. Firstly, notice that,
\begin{equation}
P_v^{\mu}+\sum_{e=1}^E \eta_{ve}q_e^{\mu}=\frac{1}{\mathcal{U}}\sum_T\bigg(P_v^{\mu}+\eta_{ve}P_{T,e}^{\mu}\bigg)\prod_{e'\notin T}\alpha_{e'}=0\ .
\end{equation}
This is because for every spanning tree we have momentum conservation at each vertex $v$ with the outgoing momentum from $v$ being $P_v^{\mu}$. Secondly, we find that,
\begin{equation}
\sum_{e=1}^E\eta_{Ie}q_e^{\mu}\alpha_e=\frac{1}{\mathcal{U}}\sum_{T_1\sqcup T_2}\bigg(\eta_{Ie}P_{T_1,T_2,e}^{\mu}\bigg)\prod_{e'\notin T_1,T_2}\alpha_{e'}\ ,
\end{equation}
where $P_{T_1,T_2,e}$ denotes the total momentum flowing from one tree to the other. The sum over all edges along the loop $I$ in the parentheses vanishes because there is net momentum flowing between $T_1$ and $T_2$ within such a loop since $T_1$ and $T_2$ are disjoint trees. Lastly, we remark that the world-line action $\mathcal{V} $ can be written in different forms among which the following is most useful. Note that,
\begin{equation}
\mathcal{V}(\alpha_e)=\sum_{e=1}^E (q_e^2-m_e^2)\alpha_e\bigg|_{\eqref{eq:LE_1},\eqref{eq:LE_2}}=\sum_{e=1}^E (q_e.q_e\alpha_e-\alpha_e m_e^2)\bigg|_{\eqref{eq:LE_1},\eqref{eq:LE_2}}\ .
\end{equation}

The above solutions are basically a theorem in combinatorics for electrical networks \cite{Bollobas:2013}. In that case the linear Landau equations are replace by Kirchhoff's current and voltage laws which are essentially of the same form \cite{Bjorken:1959, Mathews:1959}. 

One can now use the expressions \eqref{eq:edge1} and \eqref{eq:edge2} which satisfy \eqref{eq:LE_1} and \eqref{eq:LE_2} respectively. As a result we obtain the following form of the action which we were referring to,
\begin{equation}
\mathcal{V}(\alpha_e)=\sum_{\text{subsets}\ S}p_S^2\mathcal{F}_S-\sum_{e=1}^E m_e^2\alpha_e\ .
\label{eq:actionF}
\end{equation}
Here $S$ denotes the proper subsets of $n$ labels (that are the $n$ external momenta) without double counting the complements $\bar{S}:=\lbrace 1,2,\dots,n\rbrace\backslash S$. There are $2^{n-1}-1$ such subsets. We have,
\begin{equation}
p_S^{\mu}:=\sum_{a\in S}p_a^{\mu}\ ,
\end{equation}
and $\mathcal{F}_S$ is defined through
\begin{equation}
\mathcal{F}_S:=\frac{1}{\mathcal{U}}\sum_{\substack{\text{spanning}\\ \text{2-forests }F_S}}\prod_{e\notin F_S}\alpha_e\ ,
\end{equation}
where a spanning 2-forest $F_S:=T_S\sqcup T_{\bar{S}}$ is the disjoint union of trees $T_S$ ($T_{\bar{S}}$) connected to all the external momenta in $S$ ($\bar{S}$) and none from $\bar{S}$ ($S$) such that every vertex belongs to either tree. This can be achieved by cutting through exactly $L+1$ edges such that sets $S$ and $\bar{S}$ are separated and hence every $\mathcal{F}_S$ is a homogeneous function of degree 1 in the $\alpha_e$'s.

In $\S$\ref{sec:NP_main} we will use the expressions \eqref{eq:edge1}, \eqref{eq:edge2} and \eqref{eq:actionF} directly and repeatedly. We do check explicitly that the linear Landau equations are satisfied and that the final form of the action matches the definition with which we begin in each case, although we do not present those computations in the paper for the interest of saving space.

\subsection{Analytic continuation through the five steps}
As promised we will now turn our attention to the five steps via which we go from an amplitude in a particular channel to the amplitude in its crossing channel by analytically continuing the external momenta (satisfying the momentum conservation and the on-shell conditions) and deforming the integration contours to avoid singularities.

As may be obvious from figure \ref{fig:1} steps I, III and V are essentially the same operations, so showing that we can execute step I by avoiding the singularities immediately guarantees that the same can be done for steps III and V as well. Similarly the steps II and IV are same operations, so we need only show the analyticity of the amplitude during step II and the analyticity during step IV follows essentially the same arguments.

\subsubsection{Analyticity during step I:}
In this step of the analytic continuation we need to discuss analyticity in the infinitesimal neighbourhoods of the physical regions. Both the external kinematics and Schwinger parameters will be deformed at the same time to demonstrate the analytic properties.

Firstly, consider the \textbf{neighbourhoods of non singular points} i.e. the points in the physical region for which,
\begin{equation}
(q_e^2-m_e^2)\neq 0\ ,\quad\forall\ e=1,\dots, E\ .
\label{eq:S1_non_sing}
\end{equation}
Although, above the production threshold, it can still happen that $\mathcal{V}=0$ somewhere along the integration contour. Now suppose, we deform the contours by giving small phases to the integration parameters viz.,
\begin{equation}
\hat{\alpha}_e:=\alpha_e e^{i\varepsilon(q_e^2-m_e^2)}\approx \alpha_e+i\varepsilon(q_e^2-m_e^2)\alpha_e+O(\varepsilon^2)\ ,
\end{equation}
where $\varepsilon$ is some positive, small, real parameter and from \eqref{eq:S1_non_sing} we deduce that all the phases are non zero but finite. Since the edge momenta can be expressed as a function of the external kinematics and the Schwinger parameters only it will also pick up some correction due to this deformation i.e. $q_e^{\mu}\rightarrow\hat{q}_e^{\mu}=q_e^{\mu}(\hat{\alpha}_e',p_i)$. Thus we observe that the world-line action also picks up a correction to the first order in $\varepsilon$,
\begin{equation}
\begin{aligned}
&\hat{\mathcal{V}}\approx\mathcal{V}+i\varepsilon\sum_{e=1}^{E}(q_e^2-m_e^2)\alpha_e\frac{\partial\mathcal{V}}{\partial\alpha_e}\bigg|_{\hat{\alpha}=\alpha}+O(\varepsilon^2)\ ,\\
\Rightarrow\ &\hat{\mathcal{V}}\approx\mathcal{V}+i\varepsilon\sum_{e=1}^{E}(q_e^2-m_e^2)^2\alpha_e+O(\varepsilon^2)\ .
\end{aligned}
\end{equation}
This implies that at the point on the integration contour where $\mathcal{V}=0$ we can simply deform the contour away from this point to obtain $\mathcal{V}\rightarrow\hat{\mathcal{V}}\neq 0$, which implies that we do not encounter the singularity on the deformed contour. For real kinematics the coefficient of $i\varepsilon$ above is strictly positive and thus to first order in $\varepsilon$ we get Im$\hat{\mathcal{V}}>0$ and thus the correct $i\varepsilon$ prescription is implemented. To show that there exists a complex neighbourhood of this kinematic point where the $i\varepsilon$ prescription still works, we deform the external kinematics by,
\begin{equation}
p_i^{\mu}\rightarrow\tilde{p}_i^{\mu}:=p_i^{\mu}+\varepsilon^2\Delta p_i^{\mu}\ ,
\label{eq:EKD}
\end{equation}   
such that it still preserves the conservation of momentum and the on-shell conditions. The deformation is chosen to be subleading to the contour deformation so that it does not negate the corrections due to the deformation in the contour. As result of \eqref{eq:edge1} the internal momenta responds as,
\begin{equation}
q_e^{\mu}\rightarrow\tilde{q}_e^{\mu}:=q_e^{\mu}+\varepsilon^2\Delta q_e^{\mu}
\end{equation}
This change affects $\hat{\mathcal{V}}$ only at the subleading order $O(\varepsilon^2)$, hence for sufficiently small $\varepsilon$, $\hat{\mathcal{V}}$ remains to be non vanishing and Im$\hat{\mathcal{V}}$ remains to be positive. Thus we conclude that there exists a complex neighbourhood of any non-singular point, however small, where the amplitude is analytic. 

Finally, let us consider the \textbf{neighbourhoods of singular points}. From the discussions of the previous subsections it should be obvious that at the singularities, the values of the Schwinger parameters (more precisely the positive values i.e. $\alpha_e>0$) actually solves the Landau equations i.e. $\alpha_e=\alpha^*_e$. Generically the $+\alpha$-Landau surfaces are co-dimension one surfaces (or curves in the case of 4-point functions) in the physical regions called the \textit{Landau curve} \cite{Iagolnitzer:2014}. In what follows we will show the analyticity of the amplitude in a complex neighbourhood of the Landau curve corresponding to the leading Landau singularity. At the very end of this discussion we will argue why this is sufficient for the proof of analyticity in the neighbourhood of subleading Landau singularities as well. 

Now, at the singular points we have,
\begin{equation}
(q_e^*)^2=m_e^2\ ,\quad\forall\ e=1,\dots,E\ .
\label{eq:LLS}
\end{equation}
This implies $\mathcal{V}^*=\mathcal{V}(\alpha_e^*,q^*_e)=\sum_e\lbrace(q^*_e)^2-m_e^2\rbrace\alpha^*_e=0$. It should be fairly obvious that by deforming the contour as in the previous does not help us avoid the singularity, since
\begin{equation}
\hat{\alpha}^*_e=\alpha^*_e e^{i\varepsilon\left\lbrace(q_e^*)^2-m_e^2\right\rbrace}=\alpha^*_e\ ,\quad\text{(owing to \eqref{eq:LLS}).}
\end{equation}
As a result $\hat{q}^*_e=q^*_e$ and $\hat{\mathcal{V}}^*=\mathcal{V}(\hat{\alpha}_e^*,\hat{q}^*_e)=\mathcal{V}(\alpha_e^*,q^*_e)=0$ identically. We can however deform the external kinematic (satisfying momentum conservation and on-shell conditions) as in \eqref{eq:EKD},
\begin{equation}
\tilde{p}_i^{\mu}:=p_i^{\mu}+\varepsilon^2\Delta p_i^{\mu}\ ,\quad\Rightarrow\quad \tilde{\hat{q}}^{*\mu}_e:=q^{*\mu}_e+\varepsilon^2\Delta q^{*\mu}_e\ .
\end{equation}  
This results in the following corrections in the Schwinger parameters and the world-line action,
\begin{eqnarray}
\tilde{\hat{\alpha}}^*_e&\approx&\alpha^*_e+O(\varepsilon^3)\ ,\\
\tilde{\hat{\mathcal{V}}}^*&=&\mathcal{V}(\tilde{\hat{\alpha}}_e,\tilde{\hat{q}}^*_e)\ =\ \hat{\mathcal{V}}^*+\varepsilon^2\sum_{e=1}^E \Delta q^{*\mu}_e\frac{\partial\tilde{\hat{\mathcal{V}}}^*}{\partial q^{*\mu}_e}\bigg|_{\tilde{q}^*_e=q^*_e}+\dots\ \nonumber\\
\Rightarrow\ \tilde{\hat{\mathcal{V}}}^*&\approx& 2\varepsilon^2\sum_{e=1}^E\Delta q^*_e.q^*_e\alpha^*_e+O(\varepsilon^3)\ .
\end{eqnarray}
So, we see that deforming the external kinematics we actually do move away from the singularity as the world-line action and by extension its derivatives w.r.t $\alpha_e$ becomes non-vanishing. Finally to ensure analyticity of the amplitude in the neighbourhood of the Landau curve we need to guarantee that this deformation lands us with the correct $i\varepsilon$ prescription. Since $\varepsilon$ is real we need,
\begin{equation}
\text{Im}\left(\sum_{e=1}^E\Delta q^*_e.q^*_e\alpha^*_e\right)>0
\label{eq:EpsilonPres}
\end{equation}
for sufficiently small $\varepsilon$.

This implies that the singular points do have neighbourhoods of analyticity, but only when approached from specific side i.e. depending on the value of $q^*_e$ at the singularities only those deformations ($\Delta q^*_e$) are allowed such that \eqref{eq:EpsilonPres} is satisfied.

Let us now consider the simple example of a 4-point function. We take the external momenta in the lightcone coordinates
\begin{equation}
p^{\mu}=(p^+,p^-,\vec{p})\ ,\ \text{with Lorentz norm}\ p^2=p^+p^- -|\vec{p}|^2\ .
\end{equation}
Furthermore we work in a Lorentz frame where,
\begin{eqnarray}
p_1^{\mu}=(p_1^+,p_1^-,\vec{p}_1)\ ,&& p_2^{\mu}=(p_2^+,p_2^-,\vec{p}_2)\ ,\label{eq:Lframe1}\\
p_3^{\mu}=(-p_2^+,-p_2^-,\vec{p}_3)\ ,&& p_4^{\mu}=(-p_1^+,-p_1^-,\vec{p}_4)\ .\label{eq:Lframe2}
\end{eqnarray} 
Note that all the transverse components are taken to be different so as to avoid co-linear kinematics. Also the readers can convince themselves that such a Lorentz frame is always possible. Once we fix the frame to have, $p_3^{\pm}=-p_2^{\pm}$, conservation of momentum automatically fixes $p_4^{\pm}=-p_1^{\pm}$. In this frame the conservation of momentum takes the form $\sum_i\vec{p}_i=0$ and the the on-shell conditions read $M_i^2=p_i^+p_i^--|\vec{p}_i|^2$. We chose,
\begin{equation}
p_1^{\pm}>0\ ,\ p_2^{\pm}>0\ \Rightarrow\ p_4^{\pm}<0,\ p_3^{\pm}<0\ ,
\end{equation}
which corresponds to 12$\rightarrow$34 scattering. 
Finally, by the use of step I we set,
\begin{equation}
\frac{p_1^+}{p_1^-}<\frac{p_2^+}{p_2^-}\ ,
\label{eq:step1}
\end{equation}  
which ensures that the momenta of particles 2 and consequently 3 lie closer to the positive axis of the lightcone than the remaining vectors. Regardless of the deformation of the external kinematics in the intermediate stages, at the end of step one all the external kinematics are kept real i.e. they lie within the physical region.  

\subsubsection{Analyticity during step II:}
So far the arguments we have made never required that the Feynman diagrams under consideration be planar thus upto this point the discussion is general and are valid for non planar diagrams as well. We will use this fact and the results quoted until now can be straightforwardly used when discussing the non planar case.

To carry out step II we need to consider analytic continuation between future and past lightcones. They are connected through specific regions of the complexified kinematic space known as the \textit{crossing domains}. Let us continue with the 4-point example we began with in the previous sub subsection. The kinematic deformation performed in step II is simply a rotation to the other side of the lightcone for particles 2 and 3 viz.,
\begin{equation}
\tilde{p}_2^{\mu}=(zp_2^+,\frac{1}{z}p_2^-,\vec{p}_2)\ ,\quad \tilde{p}_3^{\mu}=(-zp_2^+,-\frac{1}{z}p_2^-,\vec{p}_3)\ .
\label{eq:step2deform}
\end{equation}
It is easy to check that such a deformation obeys the momentum conservation as well as the on-shell condition since,
\begin{eqnarray}
\tilde{p}_2^+\tilde{p}_2^--|\vec{\tilde{p}}_2|^2&=&z.\frac{1}{z}p_2^+p_2^--|\vec{p}_2|^2\ =\ M_2^2\\
\tilde{p}_3^+\tilde{p}_3^--|\vec{\tilde{p}}_3|^2&=&z.\frac{1}{z}(-p_2^+)(-p_2^-)-|\vec{p}_3|^2\ =\ M_3^2\ .
\end{eqnarray}
We will be using the simplest possible deformation between $z=1$ and $z=-1$ along a path in the upper-half plane, Im$z>$0 outside of the unit semi-circle, $|z|^2\geq1$.

From the form of the deformation it is obvious that only the components $\mu=\pm$ are affected so we will focus our attention to these components only. Now notice that the solution to the linear Landau equations for the internal edge momenta $q_e$'s are linear in the external momenta so in the Lorentz frame under consideration they take the following form,
\begin{equation}
q_e^{\pm}=p_1^{\pm}f_{e,14}+z^{\pm1}p_2^{\pm}f_{e,23}\ .
\end{equation} 
The coefficients $f_{e,ij}$ have the general form,
\begin{equation}
f_{e,ij}:=\frac{1}{\mathcal{U}}\sum_T\eta_{e,ij}^T\prod_{e'\notin T}\alpha_e'
\end{equation}
where $\eta_{e,ij}^T$ equals to +1 ($-$1) if the unique path from $i$ to $j$ along the spanning tree T passes through $e$ with the same (opposite) orientation and 0 otherwise. The imaginary part of the quadratic Landau equation, due to the aforementioned deformation reads,
\begin{equation}
\text{Im}(\tilde{q}_e^2-m_e^2)=\text{Im}z\left(p_2^+p_1^- - \frac{1}{|z|^2}p_1^+p_2^-\right)f_{e,14}f_{e,23}=0\ .
\label{eq:step2}
\end{equation}
Thus to show that we can avoid the singularities during this step we have to argue that \eqref{eq:step2} cannot be satisfied for all $e$ simultaneously. 

\begin{figure}[h!]
\centering
\includegraphics[scale=0.5]{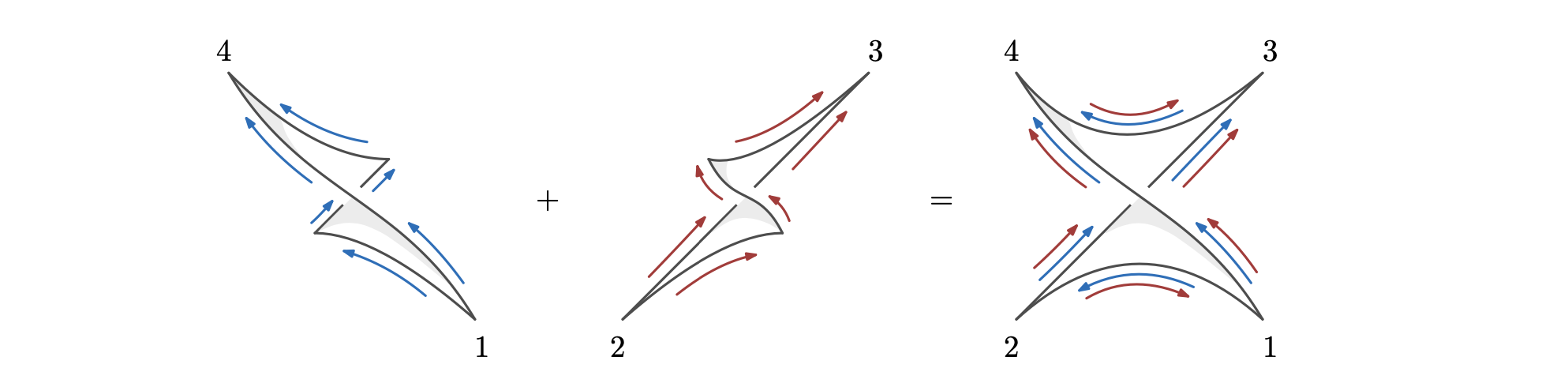}
\caption{Flow of momenta in the direction of $p_1^{\pm}$ (blue) and $p_2^{\pm}$ (red). The \textit{side} edges of the diagram always have non-zero components in both directions preventing the formation of singularities. (We acknowledge this figure to be an exact replica of figure 8 of \cite{Mizera:2021fap})}
\label{fig:2}
\end{figure}

To show this, let us note that given \eqref{eq:step1} and the fact that we move along a path in the upper half $z$ plane such that $|z|^2\geq1$ the only way \eqref{eq:step2} can be satisfied is if,
\begin{equation}
\begin{aligned}
&f_{e,14}=0\ ,\ \text{or}\ \ f_{e,23}=0\ ,\ \text{or}\ \ f_{e,14}=f_{e,23}=0\ ,\\
&\Rightarrow\ \ q_e^{\pm}\propto p_2^{\pm}\ ,\ \text{or}\ \ q_e^{\pm}\propto p_1^{\pm}\ ,\ \text{or}\ \ q_e^{\pm}=0\ .
\end{aligned}
\end{equation}
It is at this point where we invoke the \textbf{planar limit}. Note that for planar Feynman diagrams, the \textit{side} edges always carry strictly positive energy so $q_e^{\pm}=0$ case is ruled out for the \textit{side} edges from the get go. For the cases of non zero $f_{e,14}$ or $f_{e,23}$ we two problems sourced by one incoming and one outgoing momentum each. For $f_{e,14}\neq 0$ it is proportional to the energy flow of the diagram sourced by particles 1 and 4, while for $f_{e,23}\neq 0$ it is proportional to the energy flow of the diagram sourced by particles 2 and 3. A pictorial depiction of these facts are presented in figure \ref{fig:2} (see, \cite{Mizera:2021fap}). 
The full Feynman diagram sourced by two incoming and two outgoing particles thus have $f_{e,14}\neq 0$ and $f_{e,23}\neq 0$ for the \textit{side} edges. This implies that for the \textit{side} edges \eqref{eq:step2} can never be satisfied as a result of which we conclude that we don't encounter any singularities when deforming the momenta through the \textit{crossing domain} from the future (past) lightcone to the past (future) lightcone.

\subsection{Amplitudes with arbitrary number of external states}
\label{sec:General}
Generalisation to the case of higher point ($n>4$) amplitudes is pretty straight forward. We club the incoming states into two non-empty sets $A$ and $B$ and the outgoing states into two non-empty sets $C$ and $D$. One can denote by $p_S^{\mu}$ the total external momentum of the particles in the set $S$. We pick a Lorentz frame in which,
\begin{eqnarray}
p_A^{\mu}=(p_A^+,p_A^-,\vec{p}_A)\ ,&& p_A^{\mu}=(p_B^+,p_B^-,\vec{p}_B)\ ,\\
p_C^{\mu}=(-p_B^+,-p_B^-,\vec{p}_C)\ , && p_D^{\mu}=(-p_A^+,-p_A^-,\vec{p}_D)\ ,
\end{eqnarray}
The lightcone components of the external particles satisfy,
\begin{eqnarray}
p_A^{\pm}&=&\sum_{a\in A}p_a^{\pm}\ =\ -\sum_{d\in D}p_d^{\pm}>0\ ,\\
p_B^{\pm}&=&\sum_{b\in B}p_b^{\pm}\ =\ -\sum_{c\in C}p_c^{\pm}>0\ ,\\
\end{eqnarray}
with all $p_a^{\pm},p_b^{\pm}>0$ and all $p_c^{\pm},p_d^{\pm}<0$ owing to the fact that they are incoming and outgoing particles respectively. The conservation of momentum reads,
\begin{equation}
\vec{p}_A+\vec{p}_B+\vec{p}_C+\vec{p}_D=\sum_{i=1}^n\vec{p}_i=0
\end{equation}
and as usual the on-shell conditions are given by $M_i^2=p_i^+p_i^- -|\vec{p}_i|^2$. For the purpose of stating the endpoint of step one in this case we use the following ratios,
\begin{equation}
\theta_S:=\frac{p_S^+}{p_S^-}
\end{equation}
which is a measure of the angle of each momentum with the positive axis of the lightcone. The endpoint of step I is taken to be the configuration in which 
\begin{equation}
\frac{\theta_a\theta_d}{\theta_A}<\frac{\theta_b\theta_c}{\theta_B}
\end{equation} 
for all $a\in A, b\in B, c\in C$ and $d\in D$. We don't require the the individual momenta to be ordered within each set.

For step II i.e. the analytic continuation through the crossing domain we perform the rotations,
\begin{equation}
\tilde{p}_b^{\mu}=(zp_b^+,\frac{1}{z}p_b^-,\vec{p}_b)\ ,\ \ \tilde{p}_c^{\mu}=(-zp_c^+,-\frac{1}{z}p_c^-,\vec{p}_c)\ ,
\end{equation}
for all $b\in B$ and $c\in C$. It is easy to check, as in the case of 4-point function, that such a deformation does not violate either conservation of momentum or the on-shell conditions and stays within the same frame as specified above. Again the analytic continuation proceeds from $z=1$ to $z=-1$ via the upper half plane Im$z>0$ through a path lying outside the unit semicircle $|z|\geq 1$. Let us now state the imaginary part of the quadratic Landau equation necessary for a singularity in this case, 
\begin{equation}
\text{Im}(\tilde{q}_e^2-m_e^2)=\text{Im}(z)\sum_{\substack{a\in A \\ d\in D}}\sum_{\substack{b\in B \\ c\in C}}\frac{p_a^-p_b^-p_c^-p_d^-}{p_A^-p_B^-}\left(\frac{\theta_b\theta_c}{\theta_B}-\frac{1}{|z|^2}\frac{\theta_a\theta_d}{\theta_A}\right)f_{e,ad}f_{e,bc}=0,\ \ \forall\ e.
\label{eq:step2gen}
\end{equation}
The proof of analyticity in this case follows essentially the same arguments as in the 4-point case. One can show that for a \textbf{planar diagram} sourced by all the external particles we must have $f_{e,bc}\neq 0$ and $f_{e,ad}\neq 0$ for the \textit{side} edges. As a result of this \eqref{eq:step2gen} cannot be satisfied for all $e$ simultaneously and hence analyticity of the amplitude holds in the crossing domain.  

This concludes the proof of analyticity of the amplitudes during step I and step II. As mentioned earlier, the other steps viz. step III, IV and V are same as either step I or step II but with some different pair of momenta, so analyticity during these steps follow immediately from the analyticity of steps I and II. So, we see that during the analytic continuation of the Feynman diagram from the channel $AB\rightarrow CD$ to the crossing channel $B\bar{C}\rightarrow D\bar{A}$ we can avoid the Landau singularities proving crossing symmetry for \textbf{planar diagrams} in perturbative QFT.\\\\

Before ending this section let us see why it is sufficient to consider only the leading Landau singularities for the proof of crossing symmetry. Firstly note that the sub-leading Landau singularities are given by boundary saddle points of \eqref{eq:Main_Integral} as mentioned earlier. This means that for these type of singularities one or more $\alpha_e=\alpha_e^*=0$ while the remaining are nonzero. As a result we see that the sub-leading Landau singularity surfaces are \textit{higher co-dimension} surfaces i.e. \textit{co-dimension $\geq2$} and we know that in $D>2$ we can always avoid higher co-dimension surfaces\footnote{Consider a $n$-point amplitude in $D$ dimensions ($n\geq D$). There are $D$ linearly independent vectors each with $D$ components. On-shell condition for each of them along with overall momentum conservation give $D+D=2D$ number of holonomic constraints. Thus number of independent variables $=D^2-2D=D(D-2)$. For $D>2$ this is non zero and hence we have some independent direction along which we can move to avoid higher co-dimension surfaces in the kinematic space.}. In \cite{Fevola:2023kaw, Fevola:2023fzn}, it was shown that sub-leading Landau singularities of a Feynman diagram $G$ is nothing but the leading Landau singularity of a \textit{reduced diagram} $G/\gamma$ (for some connected sub-diagram $\gamma\subset G$) obtained from $G$ by contracting all the edges in $\gamma$ and identifying all the vertices in $\gamma$. Thus analytic continuation of the external momenta avoiding a sub-leading Landau singularity surface for the diagram $G$, amounts to the analytic continuation of the external momenta avoiding the leading Landau singularity surface for the diagram $G/\gamma$.

\section{Proof of crossing in the non planar case}
\label{sec:NP_main}
Now with all the ingredients at our disposal we turn to the case of non-planar diagrams. We will first go through the case of the non-planar diagrams occurring at 2-loop level and represent the situation with flow diagrams as in figure \ref{fig:2}. 

We can carry out the steps of the analytic continuation described in the previous section as follows. 
\begin{enumerate}
\item{We first go to a Lorentz' frame characterised by \eqref{eq:Lframe1} and \eqref{eq:Lframe2} in the lightcone coordinates, so that we have $p_4^{\pm}=-p_1^{\pm}$ and $p_3^{\pm}=-p_2^{\pm}$.}
\item{Then we carry out step I to reach a point in the momentum space for which \eqref{eq:step1} is satisfied.} 
\item{Finally we carry out step II using the deformation \eqref{eq:step2deform} so that for the imaginary part of the quadratic Landau equations we get,
\begin{equation}
\text{Im}(\tilde{q}_e^2-m_e^2)=\text{Im}z\left(p_2^+p_1^- - \frac{1}{|z|^2}p_1^+p_2^-\right)f_{e,14}f_{e,23}\ .
\end{equation}
where internal edges take the form $q_e^{\pm}=f_{e,14}p_1^{\pm}+f_{e,23}p_2^{\pm}$ in this Lorentz' frame.}
\end{enumerate} 
Now if \eqref{eq:step2} is satisfied i.e.
\begin{equation}
\text{Im}(\tilde{q}_e^2-m_e^2)=\text{Im}z\left(p_2^+p_1^- - \frac{1}{|z|^2}p_1^+p_2^-\right)f_{e,14}f_{e,23}=0\ ,
\end{equation}
for all $e$ then only we encounter a leading Landau singularity. This is satisfied if and only if,
\begin{equation}
f_{e,14}=0\ \text{and/or}\ f_{e,23}=0,\ \forall\ e.
\end{equation}
since \eqref{eq:step1} and the fact that we go from $z=1$ to $z=-1$ along a path for which $|z|\geq 1$ via the upper half plane i.e. $\text{Im}z\geq 0$, ensures that the factor multiplying $f_{e,14}f_{e,23}$ is non vanishing. To prove that we encounter no leading Landau singularity during step II we need to show that there exists at least one $e$ for which $f_{e,14}\neq 0$ and $f_{e,23}\neq 0$. As will become clear, for our proof we only use the fact that the leading Landau singularities that are relevant for evaluating the integral \eqref{eq:Main_Integral} lie on the $+\alpha$-Landau surface as argued in the previous section. We do not need the explicit solutions for $\alpha$ that solves the Landau equations.

\subsection{The case of 2-loop}
At 2-loop level we find only two types of non-planar diagrams. For both of them there is only one propagator which is non-planar. 

\subsubsection{Case 1:}
\begin{figure}[h!]
\centering
\includegraphics[scale=0.6]{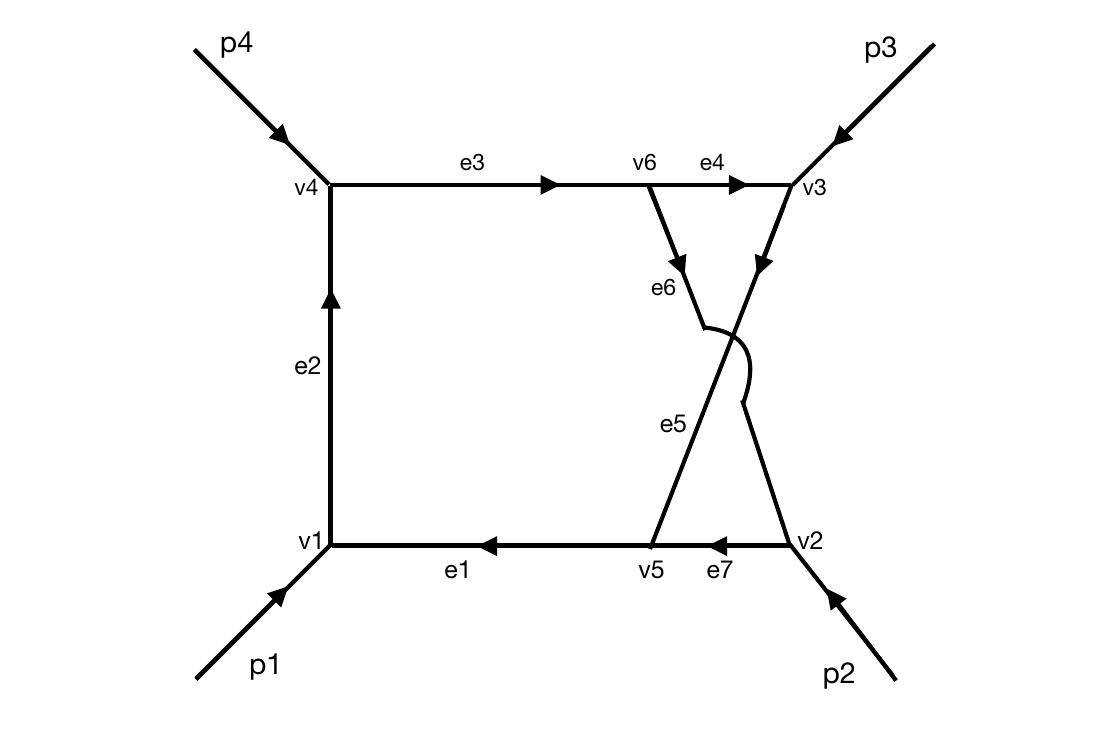}
\caption{Non planar diagram at 2-loop considered in Case 1.}
\label{fig:3}
\end{figure}

\begin{figure}[h!]
\centering
\includegraphics[scale=0.6]{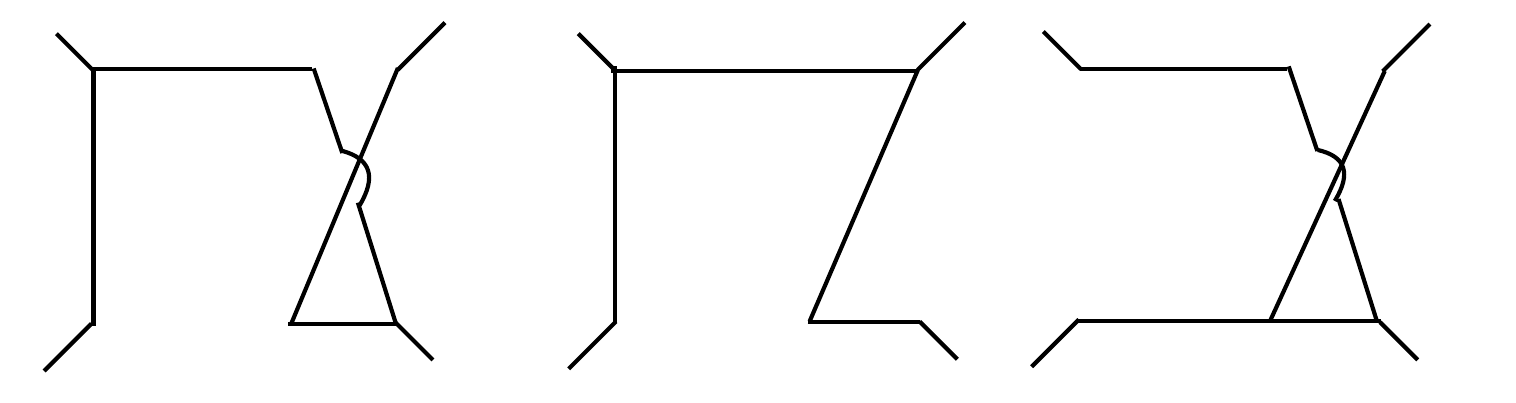}
\caption{Spanning trees corresponding to $\lbrace e_1,e_4\rbrace,\lbrace e_1,e_6\rbrace$ and $\lbrace e_2,e_4\rbrace$ respectively. The direction of flow for each internal edge is the same as the original diagram depicted in figure \ref{fig:3}.}
\label{fig:4}
\end{figure}

In the first case we have the non-planar diagram displayed in figure \ref{fig:3}. We label the internal edges in the following way without loss of generality,
\begin{equation}
\begin{aligned}
&e_1=\ell_1\ ,\ \ e_2=p_1+\ell_1\ ,\ \ e_3=p_1+p_4+\ell_1\ ,\ \ e_4= p_1+p_4+\ell_1-\ell_2\ ,\\
&e_5=p_1+p_4+p_3+\ell_1-\ell_2\ ,\ \ e_6=\ell_2\ ,\ \ e_7=p_2+\ell_2\ .
\end{aligned}
\end{equation}
We label the vertices in the counterclockwise direction so that the external momenta $p_i$ attaches to the $i$-th vertex. We give only the non zero coefficients $\eta_{ve}$ and $\eta_{Ie}$ below, whereas the rest are all zero.
\begin{equation}
\begin{aligned}
&\eta_{v_1e_1}=1,\ \eta_{v_1e_2}=-1,\ \eta_{v_2,e_6}=1,\ \eta_{v_2e_7}=-1,\ \eta_{v_3e_4}=1,\ \eta_{v_3e_5}=-1,\ \eta_{v_4e_2}=1,\ \eta_{v_4e_3}=-1,\\
&\eta_{v_5e_1}=-1,\ \eta_{v_5e_5}=1,\ \eta_{v_5e_7}=1,\ \eta_{v_6e_3}=1,\ \eta_{v_6e_4}=-1,\ \eta_{v_6e_6}=-1.\\
\\
&\eta_{\ell_1e_i}=1,\ \forall\ i=1,2,3,4,5\ ;\ \ \eta_{\ell_2e_i}=-1,\ \forall\ i=4,5\ \text{and}\ \eta_{\ell_2e_i}=1,\ \forall\ i=6,7.
\end{aligned}
\end{equation}
Finally, we need all possible spanning trees $T$ for the diagram in figure \ref{fig:3}. In what follows we denote the set of spanning trees by $\mathcal{T}^C$ for which each element provides the set of internal edges we need to drop to get the corresponding spanning tree $T$.
\begin{eqnarray}
\mathcal{T}^C=\bigg\lbrace \lbrace e_1,e_4\rbrace,\lbrace e_1,e_6\rbrace,\lbrace e_2,e_4\rbrace,\lbrace e_2,e_5\rbrace,\lbrace e_2,e_6\rbrace,\lbrace e_2,e_7\rbrace,\lbrace e_3,e_5\rbrace,\lbrace e_3,e_7\rbrace,\lbrace e_4,e_7\rbrace,\lbrace e_5,e_6\rbrace\bigg\rbrace.\nonumber\\
\end{eqnarray}
As an example we provide the first three trees corresponding to these elements pictorially in figure \ref{fig:4}. 

Using momentum conservation one can easily work out the $P_{T,e}$'s for each spanning tree $T$ and the edge $e$, setting $P_{T,e'}=0$ for $e'\notin T$. 

With all these quantities now in place, we can compute $\mathcal{U}$ using \eqref{eq:expressionU} to get,
\begin{equation}
\mathcal{U}=\alpha_3 \alpha_5 + \alpha_5 \alpha_6 + \alpha_1 (\alpha_4 + \alpha_6) + \alpha_3 \alpha_7 + \alpha_4 \alpha_7 + \alpha_2 (\alpha_4 + \alpha_5 + \alpha_6 + \alpha_7)\ .
\end{equation} 
We don't give the result for all the $q_e$'s here since the full expressions are quite large and for our purpose we don't need all of them, but the reader can work them out straightforwardly using \eqref{eq:edge1}. 

Let us first set $p_4^{\pm}=-p_1^{\pm}$ and $p_3^{\pm}=-p_2^{\pm}$ so that we are in the Lorentz' frame that we want, and then let us look at $q_{e_6}$ which is given by \footnote{Working out the internal momenta for all the edges, the readers can convince themselves that in this case, for $e=e_4,e_5,e_6,e_7$ we have $f_{e,14}\neq 0$ and $f_{e,23}\neq 0$ by the same argument. It is sufficient to consider any one of them for the proof.},
\begin{equation}
q_{e_6}^{\pm}= -\frac{\alpha_2 (\alpha_4+\alpha_5)}{\mathcal{U}}p_1^{\pm} - \frac{(\alpha_4 \alpha_7+\alpha_2 (\alpha_5+\alpha_7)+\alpha_3 (\alpha_5+\alpha_7))}{\mathcal{U}}p_2^{\pm}\ .
\end{equation}
Thus we see that,
\begin{equation}
f_{e_6,14}= -\frac{\alpha_2 (\alpha_4+\alpha_5)}{\mathcal{U}}\ \ \text{and}\ \ f_{e_6,23}= - \frac{(\alpha_4 \alpha_7+\alpha_2 (\alpha_5+\alpha_7)+\alpha_3 (\alpha_5+\alpha_7))}{\mathcal{U}}\ .
\end{equation}
Now since for the leading Landau singularities on the $+\alpha$-Landau surface we must have $\alpha_e>0,\ \forall\ e$, hence we see that for $e=e_6$ both $f_{e_6,14}\neq 0$ and $f_{e_6,23}\neq 0$. As a result, we can state that we do not encounter any singularities during step II and hence this non planar diagram has crossing symmetry. 

\subsubsection{Case 2:}
\begin{figure}[h!]
\centering
\includegraphics[scale=0.2]{"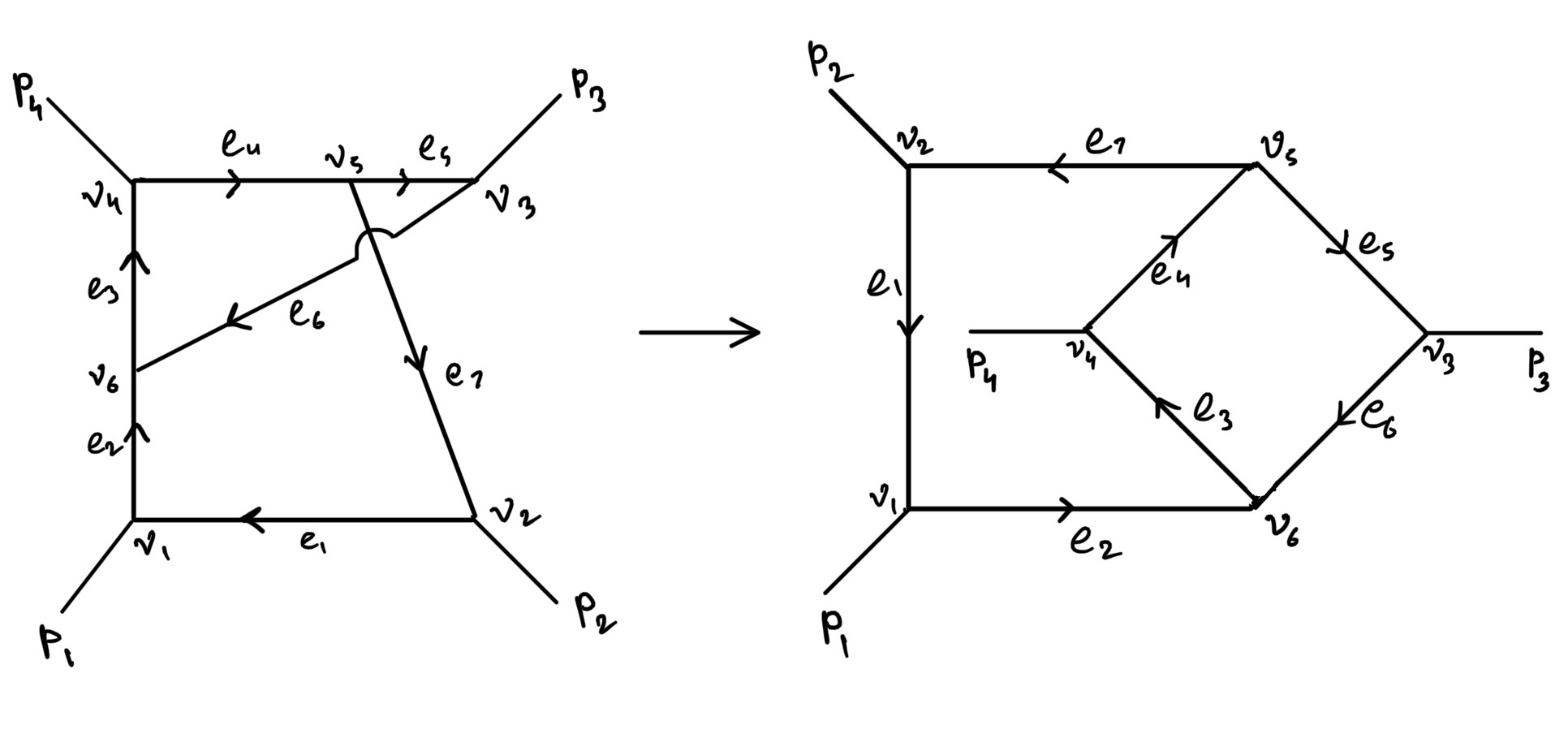"}
\caption{Non planar diagram at 2-loop considered in Case 2.}
\label{fig:5}
\end{figure}

In the second case we have the diagram depicted in figure \ref{fig:5}. We do the same exercise but now with,
\begin{equation}
\begin{aligned}
&e_1=p_2+\ell_1\ ,\ \ e_2=p_1+p_2+\ell_1\ ,\ \ e_3=p_1+p_2+\ell_1+\ell_2\ ,\\
&e_4= p_1+p_2+p_4+\ell_1+\ell_2\ ,\ \ e_5=p_1+p_2+p_4+\ell_2\ ,\ \ e_6=\ell_2\ ,\ \ e_7=\ell_1\ .
\end{aligned}
\end{equation}
The coefficients $\eta_{ve}$ and $\eta_{Ie}$ can again be worked out easily viz.,
\begin{equation}
\begin{aligned}
&\eta_{v_1e_1}=1,\ \eta_{v_1e_2}=-1,\ \eta_{v_2,e_1}=-1,\ \eta_{v_2e_7}=1,\ \eta_{v_3e_5}=1,\ \eta_{v_3e_6}=-1,\ \eta_{v_4e_3}=1,\ \eta_{v_4e_4}=-1,\\
&\eta_{v_5e_4}=1,\ \eta_{v_5e_5}=-1,\ \eta_{v_5e_7}=-1,\ \eta_{v_6e_2}=1,\ \eta_{v_6e_3}=-1,\ \eta_{v_6e_6}=1.\\
\\
&\eta_{\ell_1e_i}=1,\ \forall\ i=1,2,3,4,7\ ;\ \ \text{and}\ \ \eta_{\ell_2e_i}=1,\ \forall\ i=3,4,5,6.
\end{aligned}
\end{equation}
The rest are all zero. All that remains now is to list all the spanning trees that are generated from this diagram by dropping two edges. The resulting set is given below with each element denoting the set of edges dropped to reach the corresponding tree.
\begin{eqnarray}
\mathcal{T}^C=\bigg\lbrace \lbrace e_1,e_3\rbrace,\lbrace e_1,e_4\rbrace,\lbrace e_1,e_5\rbrace,\lbrace e_1,e_6\rbrace,\lbrace e_2,e_4\rbrace,\lbrace e_2,e_5\rbrace,\lbrace e_3,e_5\rbrace,\lbrace e_3,e_7\rbrace,\lbrace e_4,e_6\rbrace,\lbrace e_6,e_7\rbrace\bigg\rbrace.\nonumber\\
\end{eqnarray}
Again one can determine the $P_{T,e}$'s by momentum conservation and setting $P_{T,e'}=0$ for $e'\notin T$. Thus the reader can now check that in this case we get by \eqref{eq:expressionU},
\begin{equation}
\mathcal{U}=\alpha_1( \alpha_3 + \alpha_4 + \alpha_5 + \alpha_6) + \alpha_2(\alpha_4 + \alpha_5) + \alpha_3(\alpha_5+ \alpha_7) + \alpha_4 \alpha_6 + \alpha_6 \alpha_7\ .
\end{equation} 
One can again work out the momentum of the internal edges via \eqref{eq:edge1}. Again keeping in mind that for $p_4^{\pm}=-p_1^{\pm}$ and $p_3^{\pm}=-p_2^{\pm}$ we have more than one $q_e$ for which we find $f_{e,14}f_{e,23}\neq 0$, for concreteness we focus specifically on $q_{e_1}$ for which we find,
\begin{equation}
q_{e_1}=-\frac{\alpha_3 \alpha_5+\alpha_2 (\alpha_4+\alpha_5)}{\mathcal{U}}p_1^{\pm}+\frac{\alpha_6 \alpha_7+\alpha_3 (\alpha_5+\alpha_7)}{\mathcal{U}}p_2^{\pm}
\end{equation}
Thus we see that for any leading Landau singularity lying on $+\alpha$-Landau surface for which $\alpha_e=\alpha^*_e>0,\ \forall\ e$ we have,
\begin{equation}
f_{e,14}=-\frac{\alpha_3 \alpha_5+\alpha_2 (\alpha_4+\alpha_5)}{\mathcal{U}}\neq 0\ \ \text{and}\ \ f_{e,23}=\frac{\alpha_6 \alpha_7+\alpha_3 (\alpha_5+\alpha_7)}{\mathcal{U}}\neq 0\ .
\end{equation}  
As a result \eqref{eq:step2} is violated for at least one $e$ and that suffices to conclude that we can carry out step II without encountering any singularity and hence this diagram is also crossing symmetric.

\subsubsection{The energy flow diagrams}
Before ending the discussion of 2-loop non-planar diagrams, we depict the energy flow at the end of step I for the two cases elaborated above by the energy flow diagram analogous to figure \ref{fig:2}. The reason being that by looking at the flow diagrams for the 3-loop cases we will identify the scenario where carrying out the step II can potentially encounter a singularity and focus our attention primarily to that case.

\begin{figure}[h!]
\centering
\includegraphics[scale=0.25]{"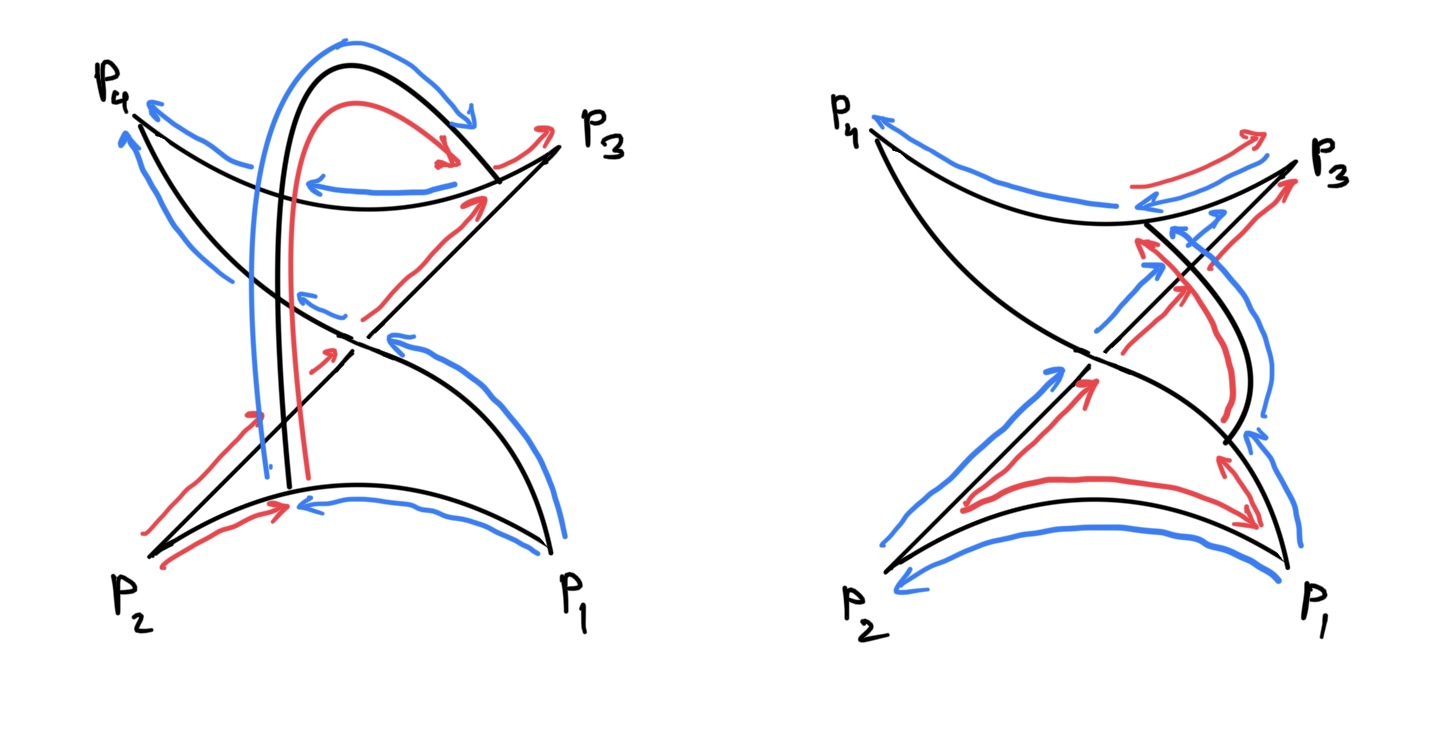"}
\caption{6(a): Flow diagram for Case 1. 6(b): Flow diagram for Case 2.}
\label{fig:6}
\end{figure}

As for the case of 2-loop non-planar diagrams, figure \ref{fig:6} depicts the energy flow from the initial states to the final states at the end of step I for both Case 1 and 2. The `Blue' arrows denote the energy flow from $p_1$ to $p_4$ for which $f_{e,14}\neq 0$, whereas the `Red' arrows denote the energy flow from $p_2$ to $p_3$ signifying $f_{e,23}\neq 0$. This means that if we find at least one internal edge carrying both `Red' and `Blue' arrows (irrespective of the direction of flow) equation \eqref{eq:step2} is violated for that edge and the proof of crossing symmetry holds true.

\subsection{The case of 3-loop}
For the case of 3-loop onwards we can divide all possible non-planar diagrams broadly into two classes viz.,
\begin{itemize}
\item{Diagrams with number of non-planar internal edges = 1.}
\item{Diagrams with number of non-planar internal edges $>$ 1.}
\end{itemize}
The first class of diagrams i.e. the ones with number of non-planar edge = 1 can be shown to be crossing symmetric straightforwardly by following the steps and arguments as in the case of 2-loop non-planar diagrams. We refer to these cases as the ``trivial" ones. We work out a couple of examples in detail in appendix \ref{app:A} to illustrate what we mean by ``trivial" in this context. 

\subsubsection{Diagrams with number of non-planar internal edges = 1 (``trivial case")} 
The results presented in this part of the section are pretty general and applies beyond the 3-loop case to non-planar diagrams at higher loops although the examples presented in the appendix \ref{app:A} illustrates the situation specifically for 3-loop non-planar diagrams. 
\begin{figure}[h!]
\centering
\includegraphics[scale=0.55]{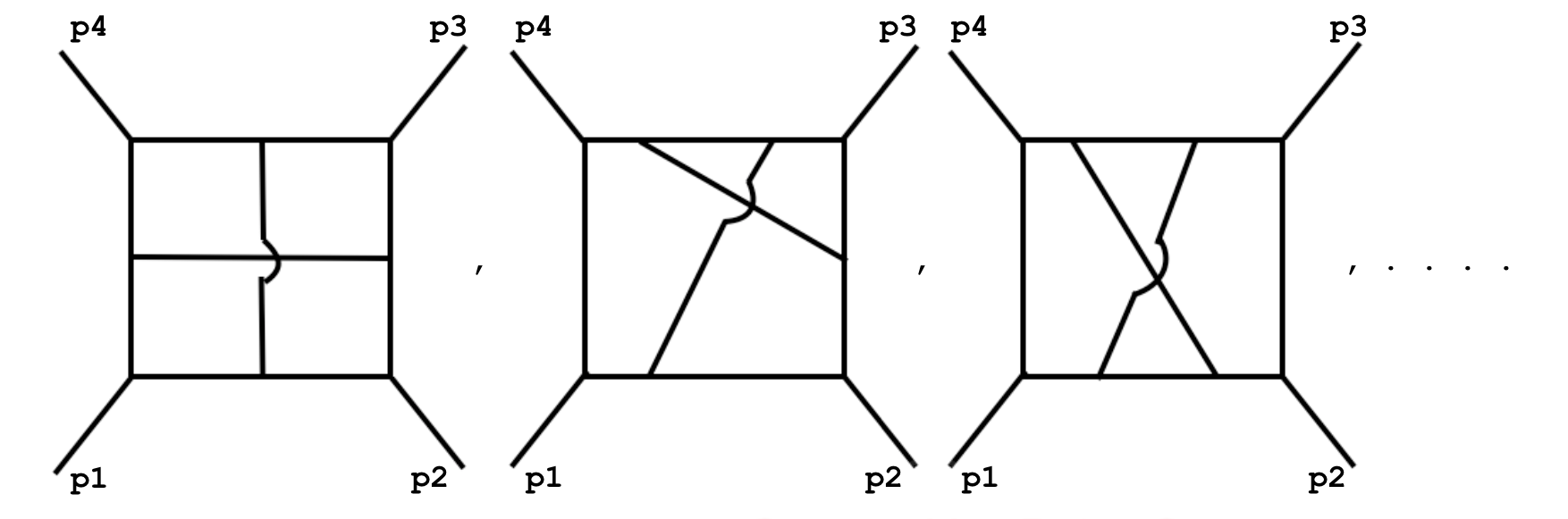}
\caption{A few examples of 3-loop diagrams with number of non-planar edges = 1}
\label{fig:7}
\end{figure}

\begin{figure}[h!]
\centering
\includegraphics[scale=0.25]{"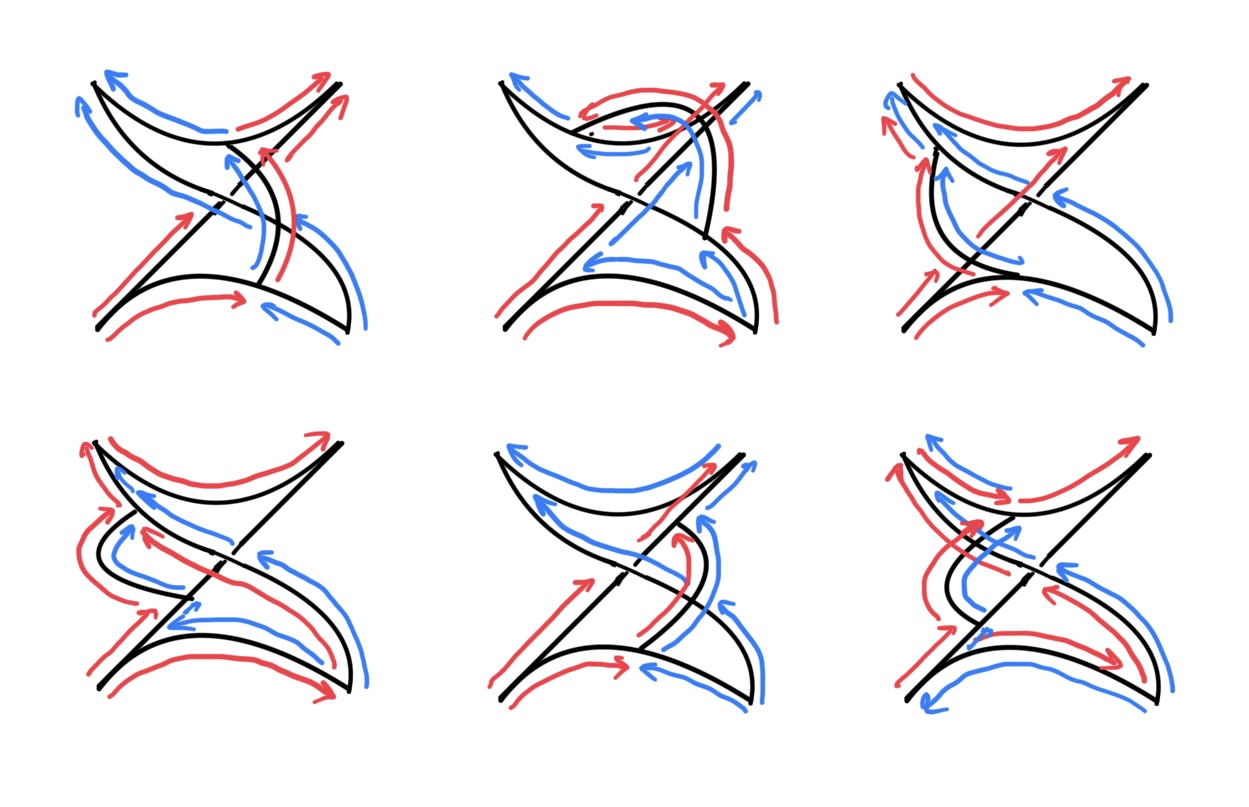"}
\caption{Flow diagrams with number of non-planar edges = 1}
\label{fig:8}
\end{figure}

In figure \ref{fig:7} we present some examples of non-planar diagrams at 3-loop order with 1 non-planar internal edge. The flow diagrams for all the ``trivial" cases are presented in figure \ref{fig:8}. From the flow of energy depicted in `Red' ($p_2$ to $p_3$) and `Blue' ($p_1$ to $p_4$) the reader can easily convince themselves that no matter which path is chosen for the energy flows, there will be at least one propagator or edge which carries both `Red' and `Blue' arrows indicating that \eqref{eq:step2} is violated for at least one edge in all these cases allowing the proof of crossing symmetry to go through for all of them. 
Note that by looking at these flow diagrams it is easy to see that even for higher loop diagrams, as long as the number of non-planar edges = 1, figure \ref{fig:8} accurately describes the energy flow in those cases. Hence we conclude that perturbative QFT amplitudes at arbitrary loop order but with number of non-planar edges = 1 are crossing symmetric.     

\subsubsection{Diagrams with number of non-planar internal edges $>$ 1 (``non-trivial cases")}
\begin{figure}[h!]
\centering
\includegraphics[scale=0.25]{"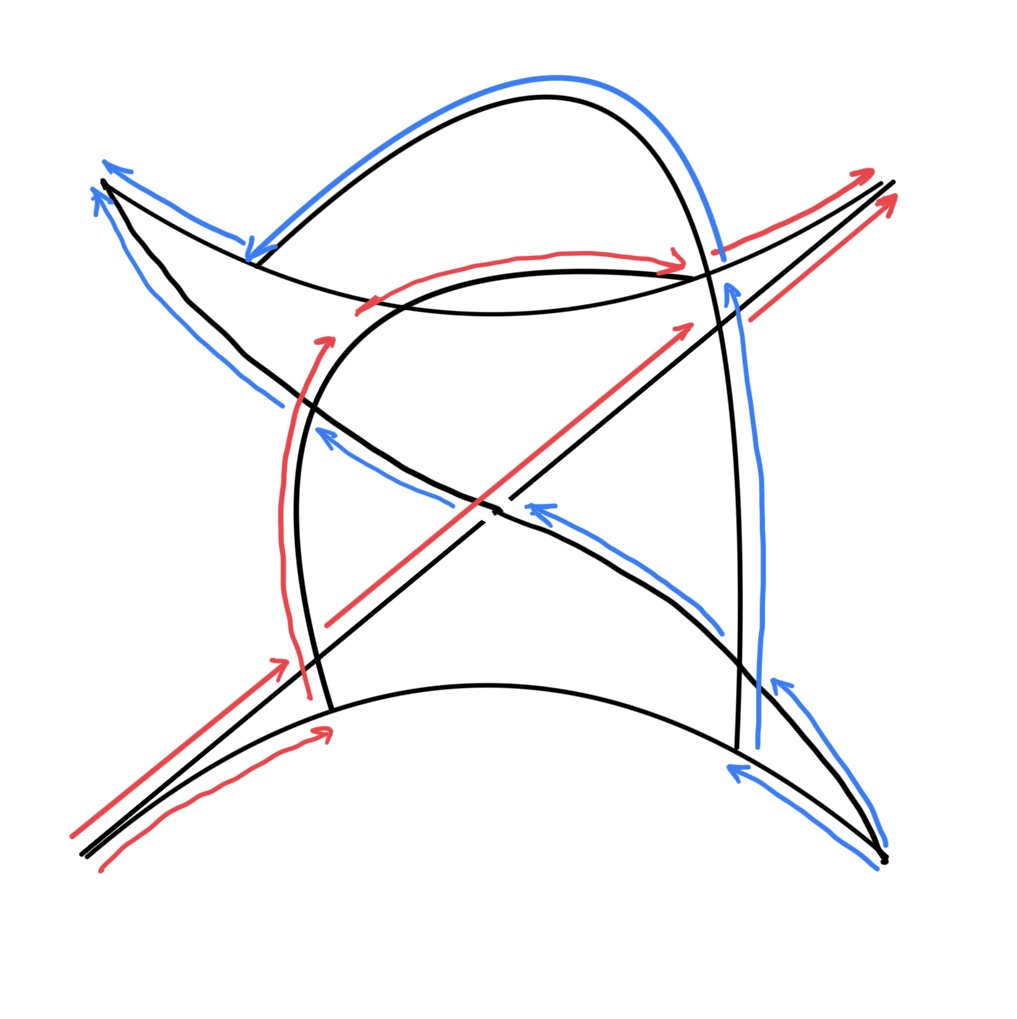"}
\caption{Flow diagram with number of non-planar edges = 2 that can potentially run into a singularity. }
\label{fig:9}
\end{figure}

Consider now the flow diagrams with more than one non-planar edge. In figure \ref{fig:9} we depict a flow diagram  with 2 non-planar edges. It is easy to convince oneself that among all possible types of flow diagrams with 2 non-planar edges, this is the only type that can potentially run into a singularity during the step II of analytic continuation.

\begin{figure}[h!]
\centering
\includegraphics[scale=0.2]{"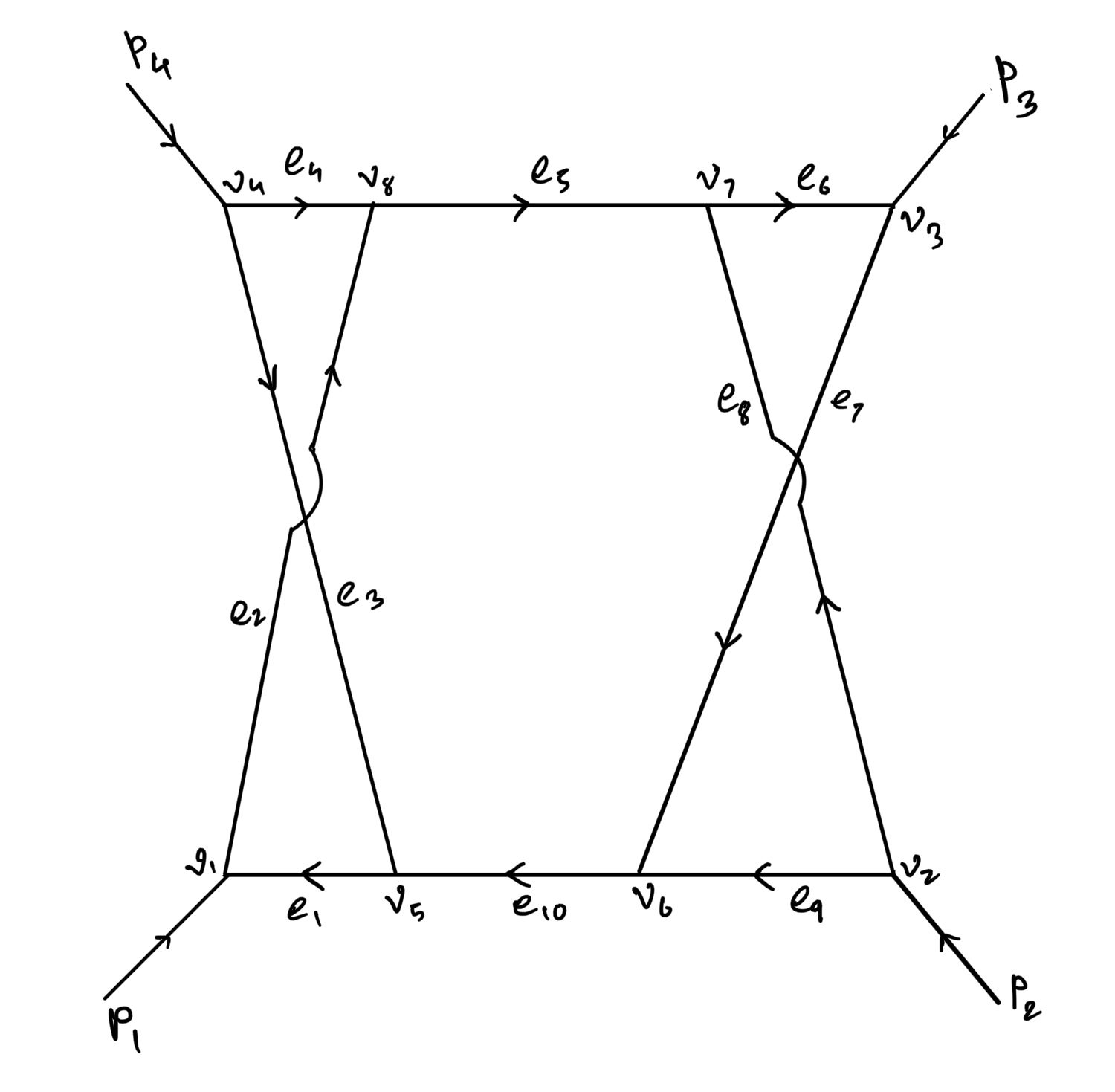"}
\caption{3-loop diagram with number of non-planar edges $>$ 1}
\label{fig:10}
\end{figure}

For the discussion in this part we restrict ourselves to the case of 3-loop non-planar diagrams. There is only one possible type at 3-loop with 2 non-planar edges depicted in figure \ref{fig:10} and it gives rise to the type showed in figure \ref{fig:9}. Let us first go through the details and determine the expression for the $q_e$'s in this case. 

We have,
\begin{equation}
\begin{aligned}
&e_1=\ell_1,\ e_2=p_1+\ell_1,\ e_3=\ell_2,\ e_4=p_4-\ell_2,\ e_5=p_1+p_4+\ell_1-\ell_2\\
&e_6=p_1+p_4+\ell_1-\ell_2+\ell_3,\ e_7=p_1+p_4+p_3+\ell_1-\ell_2+\ell_3\\
&e_8=\ell_3,\ e_9=p_2-\ell_3,\ e_{10}=\ell_1-\ell_2\ .
\end{aligned}
\end{equation}
The non zero $\eta_{ve}$ and $\eta_{Ie}$ components are given by,
\begin{equation}
\begin{aligned}
&\eta_{v_1,e_1}=1,\ \eta_{v_1,e_2}=-1,\ \eta_{v_2,e_8}=-1,\ \eta_{v_2,e_9}=-1,\ \eta_{v_3,e_6}=1,\ \eta_{v_3,e_7}=-1,\ \eta_{v_4,e_3}=-1,\\ 
&\eta_{v_4,e_4}=-1,\ \eta_{v_5,e_1}=-1,\ \eta_{v_5,e_3}=1,\ \eta_{v_5,e_10}=1,\ \eta_{v_6,e_7}=1,\ \eta_{v_6,e_9}=1,\ \eta_{v_6,e_{10}}=-1,\\
&\eta_{v_7,e_5}=1,\ \eta_{v_7,e_6}=-1,\ \eta_{v_7,e_8}=1,\ \eta_{v_8,e_2}=1,\ \eta_{v_8,e_4}=1,\ \eta_{v_8,e_5}=-1\ .\\
\\
&\eta_{\ell_1,e_i}=1,\ \forall i=1,2,5,6,7,10\ ,\ \eta_{\ell_2,e_i}=1,\ \forall i=3\ \text{and}\ \eta_{\ell_2,e_i}=-1,\ \forall i=4,5,6,7,10\ ,\\
&\eta_{\ell_3,e_i}=1,\ \forall i=6,7,8\ \text{and}\ \eta_{\ell_3,e_i}=-1,\ \forall i=9\ .
\end{aligned}
\end{equation}
Finally the set of spanning trees is given by,
\begin{equation}
\begin{aligned}
&\mathcal{T}^C=\bigg\lbrace\lbrace e_1,e_4,e_6\rbrace,\lbrace e_1,e_4,e_7\rbrace,\lbrace e_1,e_4,e_8\rbrace,\lbrace e_1,e_5,e_7\rbrace,\lbrace e_1,e_5,e_9\rbrace,\lbrace e_1,e_6,e_9\rbrace,\lbrace e_2,e_3,e_6\rbrace,\\
&\quad\lbrace e_2,e_3,e_7\rbrace,\lbrace e_2,e_3,e_8\rbrace,\lbrace e_2,e_3,e_9\rbrace,\lbrace e_2,e_6,e_9\rbrace,\lbrace e_2,e_6,e_{10}\rbrace,\lbrace e_2,e_7,e_8\rbrace,\lbrace e_2,e_8,e_{10}\rbrace,\\
&\quad\lbrace e_3,e_5,e_7\rbrace,\lbrace e_3,e_5,e_9\rbrace,\lbrace e_3,e_6,e_9\rbrace,\lbrace e_3,e_7,e_8\rbrace,\lbrace e_4,e_6,e_9\rbrace,\lbrace e_4,e_6,e_{10}\rbrace,\lbrace e_4,e_7,e_8\rbrace,\\
&\quad\lbrace e_4,e_8,e_{10}\rbrace,\lbrace e_1,e_7,e_8\rbrace,\lbrace e_1,e_4,e_9\rbrace\bigg\rbrace\ .
\end{aligned}
\end{equation}
Using these elements and equation \eqref{eq:expressionU} we get,
\begin{equation}
\begin{aligned}
&\mathcal{U}=\alpha_1 \alpha_4 \alpha_6 + \alpha_1 \alpha_4 \alpha_7 + \alpha_1 \alpha_5 \alpha_7 + \alpha_3 \alpha_5 \alpha_7 + \alpha_1 \alpha_4 \alpha_8 + \alpha_1 \alpha_7 \alpha_8 + \alpha_3 \alpha_7 \alpha_8 \\
&\quad+ \alpha_4 \alpha_7 \alpha_8 + \alpha_10 (\alpha_2 + \alpha_4) (\alpha_6 + \alpha_8) + \alpha_1 \alpha_4 \alpha_9 + \alpha_1 \alpha_5 \alpha_9 + \alpha_3 \alpha_5 \alpha_9 + \alpha_1 \alpha_6 \alpha_9\\
&\quad + \alpha_3 \alpha_6 \alpha_9 + \alpha_4 \alpha_6 \alpha_9 + 
 \alpha_2 (\alpha_7 \alpha_8 + \alpha_6 \alpha_9 + \alpha_3 (\alpha_6 + \alpha_7 + \alpha_8 + \alpha_9))
\end{aligned}
\end{equation}
and using equation \eqref{eq:edge1} we get the expressions for all the internal edge momenta $q_e$'s from which one can extract the following information,
\begin{equation}
\begin{aligned}
&f_{e_1,14}=-\frac{\alpha_4 (\alpha_7 \alpha_8+\alpha_6\alpha_9)+\alpha_{10} (\alpha_2+\alpha_4) (\alpha_6+\alpha_8)+\alpha_2 (\alpha_7 \alpha_8+\alpha_6 \alpha_9+\alpha_3 (\alpha_6+\alpha_7+\alpha_8+\alpha_9))}{\mathcal{U}}\\
&f_{e_1,23}=-\frac{(\alpha_3+\alpha_4)(\alpha_6\alpha_9-\alpha_7\alpha_8)}{\mathcal{U}}
\end{aligned}
\end{equation}
\begin{equation}
\begin{aligned}
&f_{e_2,14}=\frac{\alpha_1 (\alpha_7 \alpha_8+\alpha_6 \alpha_9+\alpha_5 (\alpha_7+\alpha_9)+\alpha_4 (\alpha_6+\alpha_7+\alpha_8+\alpha_9))+\alpha_3 (\alpha_5 (\alpha_7+\alpha_9)+\alpha_6 \alpha_9+\alpha_7 \alpha_8)}{\mathcal{U}}\\
&f_{e_2,23}=-\frac{(\alpha_3+\alpha_4)(\alpha_6\alpha_9-\alpha_7\alpha_8)}{\mathcal{U}}
\end{aligned}
\end{equation}
\begin{equation}
\begin{aligned}
&f_{e_3,14}=-\frac{(\alpha_2+\alpha_4)\alpha_7 \alpha_8+\alpha_{10} (\alpha_2+\alpha_4)(\alpha_6+\alpha_8)+(\alpha_2+\alpha_4)\alpha_6\alpha_9+\alpha_1\alpha_4 (\alpha_6+\alpha_7+\alpha_8+\alpha_9)}{\mathcal{U}}\\
&f_{e_3,23}=\frac{(\alpha_1+\alpha_2)(\alpha_6\alpha_9-\alpha_7\alpha_8)}{\mathcal{U}}
\end{aligned}
\end{equation}
\begin{equation}
\begin{aligned}
&f_{e_4,14}=-\frac{\alpha_3 (\alpha_7 \alpha_8+\alpha_6 \alpha_9+\alpha_5 (\alpha_7+\alpha_9)+\alpha_2 (\alpha_6+\alpha_7+\alpha_8+\alpha_9))+\alpha_1 (\alpha_5 (\alpha_7+\alpha_9) +\alpha_7\alpha_8+\alpha_6 \alpha_9)}{\mathcal{U}}\\
&f_{e_4,23}=-\frac{(\alpha_1+\alpha_2)(\alpha_6\alpha_9-\alpha_7\alpha_8)}{\mathcal{U}}
\end{aligned}
\end{equation}
\begin{equation}
\begin{aligned}
&f_{e_5,14}=\frac{(\alpha_1 \alpha_4-\alpha_2 \alpha_3) (\alpha_6+\alpha_7+\alpha_8+\alpha_9)}{\mathcal{U}}\\
&f_{e_5,23}=-\frac{(\alpha_1+\alpha_2+\alpha_3+\alpha_4)(\alpha_6\alpha_9-\alpha_7\alpha_8)}{\mathcal{U}}
\end{aligned}
\end{equation}
\begin{equation}
\begin{aligned}
&f_{e_6,14}=\frac{(\alpha_1\alpha_4-\alpha_2\alpha_3) (\alpha_8+\alpha_9)}{\mathcal{U}}\\
&f_{e_6,23}=\frac{\alpha_7(\alpha_3 \alpha_5 +\alpha_8(\alpha_1+\alpha_2+\alpha_3+\alpha_4))+\alpha_3(\alpha_5\alpha_9+\alpha_2 (\alpha_7+\alpha_9))+\alpha_1(\alpha_4 (\alpha_7+\alpha_9)+\alpha_5 (\alpha_7+\alpha_9))}{\mathcal{U}}
\end{aligned}
\end{equation}
\begin{equation}
\begin{aligned}
&f_{e_7,14}=\frac{(\alpha_1\alpha_4-\alpha_2\alpha_3) (\alpha_8+\alpha_9)}{\mathcal{U}}\\
&f_{e_7,23}=-\frac{\alpha_1\alpha_4(\alpha_6+\alpha_8)+\alpha_{10} (\alpha_2+\alpha_4) (\alpha_6+\alpha_8)+\alpha_6\alpha_9(\alpha_1+\alpha_3+\alpha_4)+\alpha_2 (\alpha_3 (\alpha_6+\alpha_8)+\alpha_6\alpha_9)}{\mathcal{U}}
\end{aligned}
\end{equation}
\begin{equation}
\begin{aligned}
&f_{e_8,14}=-\frac{(\alpha_1\alpha_4-\alpha_2\alpha_3) (\alpha_6+\alpha_7)}{\mathcal{U}}\\
&f_{e_8,23}=\frac{\alpha_6\alpha_9(\alpha_1+\alpha_2+\alpha_3+\alpha_4)+(\alpha_7+\alpha_9)(\alpha_3\alpha_5+\alpha_2\alpha_3+\alpha_1\alpha_4+\alpha_1\alpha_5)}{\mathcal{U}}
\end{aligned}
\end{equation}
\begin{equation}
\begin{aligned}
&f_{e_9,14}=\frac{(\alpha_1\alpha_4-\alpha_2\alpha_3) (\alpha_6+\alpha_7)}{\mathcal{U}}\\
&f_{e_9,23}=\frac{\alpha_1\alpha_4(\alpha_6+\alpha_8)+\alpha_7\alpha_8(\alpha_1+\alpha_2+\alpha_3+\alpha_4)+(\alpha_2\alpha_3+\alpha_{10}(\alpha_2+\alpha_4)) (\alpha_6+\alpha_8)}{\mathcal{U}}
\end{aligned}
\end{equation}
\begin{equation}
\begin{aligned}
&f_{e_{10},14}=f_{e_5,14}\\
&f_{e_{10},23}=f_{e_5,23}
\end{aligned}
\end{equation}
As one can point out easily, it is not possible in this case to argue just from the fact that $\alpha_e>0, \forall e$ (as in $+\alpha$-Landau surface) that there will be at least one internal edge that has $f_{e,14}\neq 0$ and $f_{e,23}\neq 0$. Indeed we have to dig a little deeper for a satisfactory answer. But note that there are only two combinations viz.,
\begin{equation}
\alpha_1\alpha_4-\alpha_2\alpha_3\quad \text{and}\quad \alpha_6\alpha_9-\alpha_7\alpha_8
\end{equation}
which can potentially vanish for $\alpha_e>0,\ \forall e$. All we need for our proof is that they cannot be 0 simultaneously on the leading $+\alpha$-Landau surface.

In this context we need at least some information about the leading Landau singularity surface. For this we use the package SOFIA \cite{Correia:2025yao} developed by Correia, Giroux and Mizera which identifies all singularity hyper-surfaces in the kinematic space for a given Feynman diagram by the use of \textit{Baikov representation} of the Feynman integrals \cite{Baikov:1996iu, Frellesvig:2024ymq}. As mentioned, this method and thereby the package identifies the leading as well as all the sub-leading singularity hyper-surfaces of any Feynman diagram. For our purpose we are interested only in the leading singularity hyper-surface. For simplicity we take all the external particles to have the same mass squared i.e. $p_i^2=M^2,\ \forall i=1,\dots,4$, as well as all the internal particles to have the same mass squared i.e. $q_e^2=m^2,\ \forall e=e_1,\dots,e_{10}$. Then the Feynman diagram depicted in figure \ref{fig:10} has all its leading Landau singularities lying on the following hyper-surface of the kinematic space.
\begin{equation}
\begin{aligned}
&\mathcal{F}=64 m^2\big(M^8 - 256 M^6 s + 320 M^4 s^2 - 128 M^2 s^3 + 
 16 s^4 - 112 M^6 t + 192 M^4 s t  - 160 M^2 s^2 t \\
 &\quad+ 32 s^3 t + 60 M^4 t^2 - 48 M^2 s t^2 + 20 s^2 t^2  - 13 M^2 t^3 + 4 s t^3+ t^4\big) - s t^4 + 64 M^6 s t + 12 M^2 s t^3\\ 
 &\quad - 5 s^2 t^3 - 80 M^4 s^2 t + 32 M^2 s^3 t - 4 s^4 t - 48 M^4 s t^2 + 40 M^2 s^2 t^2 - 8 s^3 t^2\ =\ 0\ .
\end{aligned}
\label{eq:sing_HS}
\end{equation}
Here $s=(p_1+p_2)^2$ and $t=(p_2+p_3)^2$ denote the Mandelstam variables satisfying,
$$s+t+u=4M^2\ ,\quad\text{with }\ u=(p_1+p_3)^2\ .$$
On the other hand the world-line action is given by,
\begin{equation}
\mathcal{V}(\alpha_e)=\frac{\mathcal{F}}{\mathcal{U}}\ ,\quad\text{where both $\mathcal{F}$ and $\mathcal{U}$ are polynomials of the $\alpha_e$'s.}
\end{equation}
For concreteness $\mathcal{F}$ is given in terms of the $\alpha_e$'s by,
\begin{equation}
\begin{aligned}
&\mathcal{F}=(\alpha_2 \alpha_3 + 
    \alpha_1\alpha_4)\alpha_6\alpha_7 M^2 + (\alpha_2\alpha_3 +\alpha_1\alpha_4)\alpha_8\alpha_9 M^2 + 
\alpha_1\alpha_2 (\alpha_7\alpha_8 + \alpha_6\alpha_9) M^2\\
&\quad + \alpha_3\alpha_4 (\alpha_7\alpha_8 + \alpha_6\alpha_9) M^2 + (\alpha_2\alpha_3\alpha_7\alpha_8 + 
   \alpha_1\alpha_4\alpha_6\alpha_9) s + (\alpha_1\alpha_4\alpha_7\alpha_8 + \alpha_2\alpha_3\alpha_6\alpha_9) (4 M^2 - s - t)\\
&\quad - (\alpha_1 + \alpha_{10} + \alpha_2 + \alpha_3 + \alpha_4 + \alpha_5 + \alpha_6 + \alpha_7 + \alpha_8 + \alpha_9) [\alpha_1 \alpha_4 \alpha_6 + 
     \alpha_1 \alpha_4 \alpha_7 + \alpha_1 \alpha_5 \alpha_7 \\
&\quad+ \alpha_3 \alpha_5 \alpha_7 + \alpha_1 \alpha_4 \alpha_8 + \alpha_1 \alpha_7 \alpha_8 + 
     \alpha_3 \alpha_7 \alpha_8 + \alpha_4 \alpha_7 \alpha_8 + \alpha_{10} (\alpha_2 + \alpha_4) (\alpha_6 + \alpha_8) + \alpha_1 \alpha_4 \alpha_9 \\
&\quad + \alpha_1 \alpha_5 \alpha_9 + \alpha_3 \alpha_5 \alpha_9 + \alpha_1 \alpha_6 \alpha_9 + \alpha_3 \alpha_6 \alpha_9 + \alpha_4 \alpha_6 \alpha_9\\
&\quad+ \alpha_2 \lbrace\alpha_7 \alpha_8 + \alpha_6 \alpha_9 + \alpha_3 (\alpha_6 + \alpha_7 + \alpha_8 + \alpha_9)\rbrace] m^2 
\end{aligned}
\label{eq:F_alpha}
\end{equation}
Looking at the symmetry of the diagram in figure \ref{fig:10} we now set,
\begin{equation}
\alpha_6=\alpha_1,\ \alpha_7=\alpha_2,\ \alpha_8=\alpha_3,\ \alpha_9=\alpha_4,\ \alpha_{10}=\alpha_5.
\label{eq:symm_diag}
\end{equation}
This implies that $\alpha_6\alpha_9-\alpha_7\alpha_8=\alpha_1\alpha_4-\alpha_2\alpha_3$.
As a check that there indeed can be a solution with this symmetry, the reader can verify that,
\begin{equation}
\frac{\partial\mathcal{F}}{\partial\alpha_{5+e}}\bigg|_{\eqref{eq:symm_diag}}=\frac{\partial\mathcal{F}}{\partial\alpha_e}\bigg|_{\eqref{eq:symm_diag}}\ \quad\forall e=1,\dots,5\ .
\end{equation}
Since the quadratic Landau equation implies,
\begin{equation}
\mathcal{V}(\alpha_e)=\frac{\mathcal{F}}{\mathcal{U}}=0,\ \text{and}\ \frac{\partial\mathcal{V}}{\partial\alpha_e}=0=\frac{1}{\mathcal{U}}\frac{\partial\mathcal{F}}{\partial\alpha_e}-\frac{\mathcal{F}}{\mathcal{U}^2}\frac{\partial\mathcal{U}}{\partial\alpha_e}\ \Rightarrow\ \mathcal{F}=0\ \text{and}\ \frac{\partial\mathcal{F}}{\partial\alpha_e}=0,\ \forall e\ ,
\end{equation}
as $\mathcal{U}$ is finite and non zero. Thus subject to \eqref{eq:symm_diag} we can rewrite \eqref{eq:F_alpha} as,
\begin{equation}
\begin{aligned}
&\mathcal{F}=2 \alpha_1\alpha_2 (\alpha_2\alpha_3 +\alpha_1\alpha_4) M^2 + 
 2\alpha_3\alpha_4 (\alpha_2\alpha_3 + \alpha_1\alpha_4) M^2 + (\alpha_2^2\alpha_3^2 +\alpha_1^2\alpha_4^2) s\\&\quad + 2\alpha_1\alpha_2\alpha_3\alpha_4 (4 MM - s - t) - 4 (\alpha_1 + \alpha_2 + \alpha_3 + \alpha_4 + \alpha_5) [\alpha_1^2\alpha_4\\
&\quad + \alpha_1 (\alpha_2 + \alpha_4) (\alpha_3 + \alpha_4 + \alpha_5) + \alpha_3 \lbrace\alpha_2^2 + \alpha_4\alpha_5 + \alpha_2 (\alpha_3 + \alpha_4 + \alpha_5)\rbrace] m^2
\end{aligned}
\label{eq:F_alphaFinal}
\end{equation}
Now looking at \eqref{eq:sing_HS} and \eqref{eq:F_alphaFinal} it is reasonable to take the following ansatz.
\begin{eqnarray}
\alpha_1&=&c_1M^2 + d_1s + e_1t\ ,\\
\alpha_2&=&c_2M^2 + d_2s + e_2t\ ,\\
\alpha_3&=&c_3M^2 + d_3s + e_3t\ ,\\
\alpha_4&=&c_4M^2 + d_4s + e_4t\ ,\\
\alpha_5&=&c_5M^2 + d_5s + e_5t\ .
\end{eqnarray}
But since we want to look for solutions that satisfy $\alpha_1\alpha_4-\alpha_2\alpha_3=0$, we must have,
\begin{equation}
c_1=rc_2,\ c_3=rc_4,\ d_1=rd_2,\ d_3=rd_4,\ e_1=re_2\ \text{and}\ e_3=re_4,\ \text{with $r\in\mathbb{R}$.}
\label{eq:sing_alpha}
\end{equation}
We put this ansatz for the $\alpha$'s back in \eqref{eq:F_alphaFinal} along with \eqref{eq:sing_alpha} to get an expression in terms of different powers of $M^2,\ s,\ t$ and $m^2$ which we can now equate with \eqref{eq:sing_HS}. For example, from the coefficient of $M^{10}$, $M^2s^4$ and $t^5$ we obtain,
\begin{eqnarray}
4 c_2^3 c_4 r^2 + 8 c_2^2 c_4^2 r^2 + 4 c_2 c_4^3 r^2&=&0\ ,\\
4 d_2^3 d_4 r^2 + 8 d_2^2 d_4^2 r^2 + 4 d_2 d_4^3 r^2&=&0\ ,\\
2 e_2^2 e_4^2 r^2&=&0\ .
\end{eqnarray}
Assuming that $r\neq0$, these equations imply that, $c_4=-c_2$ or $c_2=0$ or $c_4=0$, $d_4=-d_2$ or $d_2=0$ or $d_4=0$ and $e_2=0$ or $e_4=0$. In a similar fashion if we try and solve for all the vanishing coefficients of the terms without an $m^2$ factor, one can verify that we are left with the following conclusion.
\begin{equation}
\text{Either}\ c_2=0,\ d_2=0,\ e_2=0,\quad\text{or}\quad\ c_4=0,\ d_4=0,\ e_4=0.
\end{equation}
This means that either $\alpha_2=0$ or $\alpha_4=0$. Hence we can conclude that if there is such a singular point for which $\alpha_1\alpha_4=\alpha_2\alpha_3$ then it has to be a sub-leading Landau singularity \textit{ergo} for any leading Landau singularity $\alpha_1\alpha_4\neq\alpha_2\alpha_3$. This in turn implies that we do find an $e$ for which both $f_{e,14}\neq0$ and $f_{e,23}\neq0$ (for example $e=e_1,\ e_2$ etc.). So we have proven that the diagram in figure \ref{fig:10} is crossing symmetric even though the energy flow diagram may indicate otherwise.

\subsection{Generalising the argument for the ``non trivial" cases}
Looking at the specific example of the ``non-trivial" case depicted above we can give a general argument as to how the proof of crossing symmetry manifests for such non-trivial cases.

So, let us consider now a diagram with arbitrary number of non planar edges ($\geq2$) at arbitrary loop order. The leading Landau singularity surface is given by,
\begin{equation}
\mathcal{V}(\alpha,p)=0,\quad\text{such that $\alpha_e\neq0,$ $\forall e$}.
\end{equation}
This is a co-dimension 1 surface. Now, when we solve for the $q_e$'s using \eqref{eq:edge1} and \eqref{eq:edge2} we will get some result which in the Lorentz frame \eqref{eq:Lframe1},\eqref{eq:Lframe2} takes the form,
\begin{equation}
q_e^{\pm}=f_{e,14}\ p_1^{\pm}+f_{e,23}\ p_2^{\pm},
\end{equation}
for the light cone components. Now, for the ``non-trivial cases", $f_{e,14}$ and $f_{e,23}$ will have some combination of the $\alpha_e$'s (some homogeneous polynomial form multiplied by an overall factor of $1/\mathcal{U}$) which may potentially vanish. Let us call these combinations, $P_1(\alpha),P_2(\alpha),\dots$. It could happen that at specific kinematic points some or all of these combinations vanish, but that is not a problem since we can always avoid isolated points in the kinematic space. The potential obstruction to proving crossing symmetry occurs when some or all of them vanish identically i.e.
\begin{equation}
P_1(\alpha)=0,\quad P_2(\alpha)=0,\quad P_3(\alpha)=0,\quad \dots\ ,
\end{equation}
such that we get either $f_{e,14}=0$ or/and $f_{e,23}=0$ for all $e$. Hence we see that the problematic surfaces are given by,
\begin{equation}
(\mathcal{V}(\alpha,p)=0)\cap (P_1(\alpha)=0)\cap (P_2(\alpha)=0)\cap (P_3(\alpha)=0)\cap\dots\ .
\end{equation} 
It is immediately evident that these are surfaces of higher co-dimension i.e. co-dimension $\geq2$, in the kinematic space. This is exactly what we observed for the diagram depicted in figure \ref{fig:10}, as we know that sub-leading Landau singularities are indeed higher co-dimension surfaces in the kinematic space. 

Note that since we can always avoid higher co-dimension surfaces in $D>2$, it will thus be possible in $D>2$ to avoid these problematic surfaces and hence there will be no obstruction to proving crossing symmetry for these ``non-trivial" cases as well.\\\\

Finally for the generalisation of these results to the case of $n$-point amplitudes (with $n>4$) can be done using $\S$\ref{sec:General}. We divide all the $n$ particles into four sets $A,B,C,D$. By convention we take all momentum to be ingoing for the Feynman diagrams so that particles with energy $E=p^0>0$ are labeled as the incoming particles and anti-particles while particles with energy $E=p^0<0$ are labelled outgoing particles and anti-particles. The sets $A$ and $B$ are for the incoming while $C$ and $D$ are for the outgoing particles. This way of dividing up the particles into four sets has the advantage of treating the process as a 4-point amplitude but now with four beams of particles as opposed to four individual ones. For the details of how to carry out the step I and step II the readers are referred to $\S$\ref{sec:General} but the proof that $f_{e,ad}\neq0$ and $f_{e,bc}\neq0$ in \eqref{eq:step2gen} for at least one $e$ in the non-planar diagrams follows the exact same arguments as presented in this section.

\section{Conclusion}
\label{sec:Conclusion}
In this work we have presented concrete examples of non-planar diagrams in perturbative QFT and showed that these diagrams are indeed crossing symmetric i.e. we can start from each of these diagrams in a given channel and then analytically continue the result of their corresponding Feynman integrals to a crossed channel via five steps depicted in figure \ref{fig:1}. This implies that the result of a given amplitude and that of the amplitude in a crossed channel is given by two different limits of the same analytic function. The symbolic representation of this fact is given by the relation
\begin{equation}
\mathcal{T}_{AB\rightarrow CD}\equiv \mathcal{T}_{B\bar{C}\rightarrow D\bar{A}}\ .
\end{equation}
For concreteness of the proof we work with 4-point amplitudes and then generalise to the case of amplitudes with arbitrary number of external states using the facts and relations presented in $\S$\ref{sec:General}.

For the cases where we have only 1 non-planar internal edge, we have argued how the proof of crossing works out at arbitrary loop order. We refer to these as the ``trivial cases". Proof of crossing in the case of more than one non-planar internal edges, is a bit more subtle. We identify the potentially problematic or ``non-trivial cases" and give the concrete proof for the case of 3-loop amplitude and based on this proof we give a general argument of proving crossing symmetry for diagrams with arbitrary number of non-planar edges at arbitrary loop order. 

Finally let us end with a few interesting future directions to pursue from this point on. An immediate question one might ask as to, does this method generalise to the case of perturbative string theory in the language of SFT where one has a Feynman diagramatic approach to string perturbation theory. The difficulty arises primarily from the fact that in SFT the $\ell^0$ components of the loop momenta variables run from $-i\infty$ to $+i\infty$ (due to non locality), so integrating out the loop momenta variables needs a more careful analysis. In the work \cite{DeLacroix:2018arq} it was shown that at least for off-shell Green's function this integration can be carried out avoiding all pinch singularities. This gives us some indication that integrating out the loop momenta variables indeed maybe possible, but one has to address all subtleties in the case on-shell amplitudes. Furthermore, one may ask as to how the methods applied in this article can be connected to the approach of axiomatic approach of Bros, Epstein, Glaser which is applicable for the case of non perturbative QFTs as well. 

\section*{Acknowledgements}
I would like to thank Alok Laddha and Amlan Chakraborty for many helpful inputs and discussions which greatly improved my understanding of the subject. I would also like to give a general thanks to Miguel Correia, Mathieu Giroux and Sebastian Mizera for making their package SOFIA available for public use and would like to thank Sebastian Mizera in particular for helpful discussions regarding its output. Finally I thank ANRF for funding my position through Project no. SERB/PHY/2023-2024/65. Initial part of this work was supported by NSF grant PHY-2310223.

\appendix
\section{Couple of examples for 3-loop diagrams with number of non-planar edges = 1}
\label{app:A}
\begin{figure}[h!]
\centering
\includegraphics[scale=0.55]{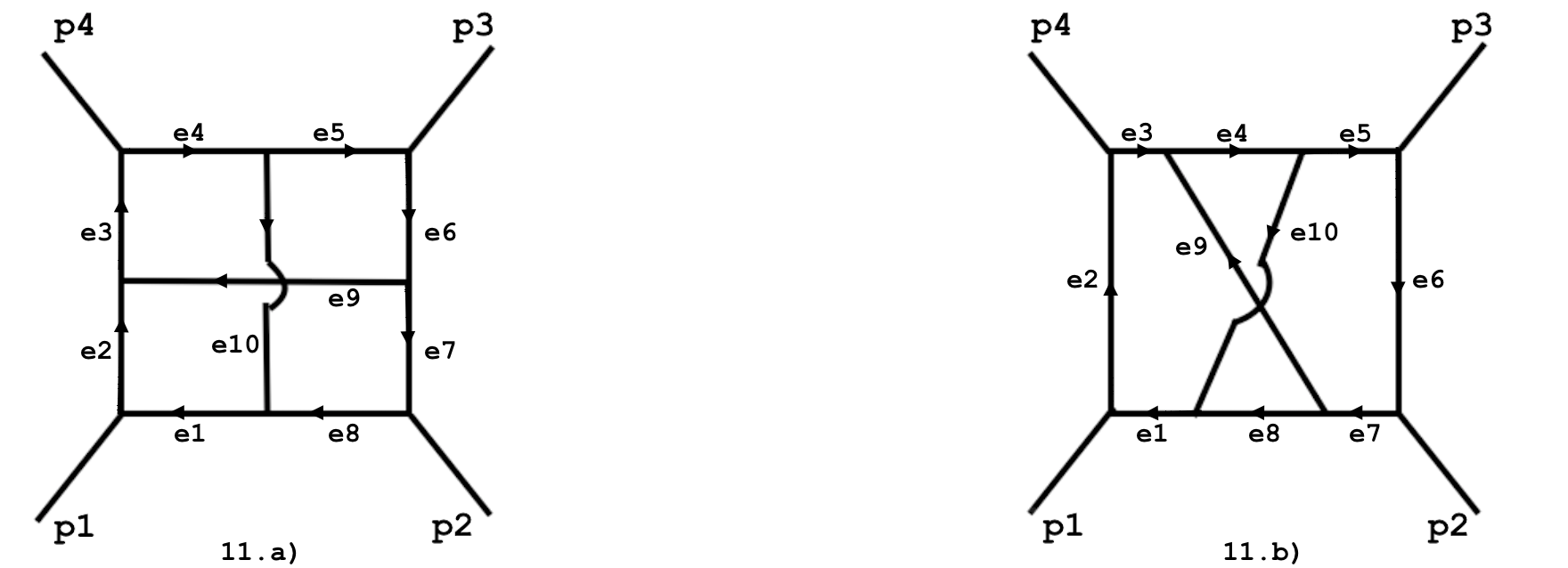}
\caption{A pair of 3-loop diagrams with 1 non-planar edge.}
\label{fig:11}
\end{figure}

The two diagrams that we consider for example in this appendix are displayed in figure \ref{fig:11}\footnote{Overall momentum conservation implies $p_1+p_2+p_3+p_4=0$}. From here on out we will refer to these diagrams as 11.a) and 11.b). As will become clear in both cases along with any other type of 3-loop diagram with 1 non-planar edge, without any further subtlety we can conclude, just from the fact that $\alpha_e>0,\ \forall e$ (on $+\alpha$-Landau surfaces) that there will be at least one internal edge for which both $f_{e,14}\neq0$ and $f_{e,23}\neq0$ just like in the case of 2-loop non-planar diagrams. It is this absence of further subtlety is why we call these cases as ``trivial".
\subsection{Diagram 11.a)}
We have the following information for this diagram.
\begin{equation}
\begin{aligned}
&e_1=\ell_1,\ e_2=p_1+\ell_1,\ e_3=p_1+\ell_1+\ell_2,\ e_4=p_1+p_4+\ell_1+\ell_2,\ e_5=p_1+p_4+\ell_1+\ell_2-\ell_3,\\
&e_6=p_1+p_4+p_3+\ell_1+\ell_2-\ell_3,\ e_7=p_1+p_4+p_3+\ell_1-\ell_3,\ e_8=\ell_1-\ell_3,\ e_9=\ell_2,\ e_{10}=\ell_3.
\end{aligned}
\end{equation}
The non zero $\eta_{ve}$ and $\eta_{Ie}$ components are given by,
\begin{equation}
\begin{aligned}
&\eta_{v_1,e_1}=1,\ \eta_{v_1,e_2}=-1,\ \eta_{v_2,e_7}=1,\ \eta_{v_2,e_8}=-1,\ \eta_{v_3,e_5}=1,\ \eta_{v_3,e_6}=-1,\ \eta_{v_4,e_3}=1,\\ 
&\eta_{v_4,e_4}=-1,\ \eta_{v_5,e_1}=-1,\ \eta_{v_5,e_8}=1,\ \eta_{v_5,e_{10}}=1,\ \eta_{v_6,e_6}=1,\ \eta_{v_6,e_7}=-1,\ \eta_{v_6,e_{9}}=-1,\\
&\eta_{v_7,e_4}=1,\ \eta_{v_7,e_5}=-1,\ \eta_{v_7,e_{10}}=-1,\ \eta_{v_8,e_2}=1,\ \eta_{v_8,e_3}=-1,\ \eta_{v_8,e_9}=1\ .\\
\\
&\eta_{\ell_1,e_i}=1,\ \forall i=1,2,3,4,5,6,7,8\ ,\ \eta_{\ell_2,e_i}=1,\ \forall i=3,4,5,6,9\ ,\\
&\eta_{\ell_3,e_i}=1,\ \forall i=10\ \text{and}\ \eta_{\ell_3,e_i}=-1,\ \forall i=5,6,7,8\ .
\end{aligned}
\end{equation}
The spanning trees are,
\begin{equation}
\begin{aligned}
&\mathcal{T}^C=\bigg\lbrace\lbrace e_2,e_4,e_6\rbrace,\lbrace e_3,e_5,e_7\rbrace,\lbrace e_2,e_5,e_7\rbrace,\lbrace e_2,e_4,e_7\rbrace,\lbrace e_1,e_4,e_6\rbrace,\lbrace e_4,e_6,e_8\rbrace,\lbrace e_1,e_5,e_7\rbrace,\\
&\quad\lbrace e_1,e_4,e_7\rbrace,\lbrace e_1,e_3,e_6\rbrace,\lbrace e_3,e_6,e_8\rbrace,\lbrace e_2,e_6,e_8\rbrace,\lbrace e_1,e_3,e_{7}\rbrace,\lbrace e_1,e_3,e_5\rbrace,\lbrace e_3,e_5,e_{8}\rbrace,\\
&\quad\lbrace e_2,e_5,e_8\rbrace,\lbrace e_2,e_4,e_8\rbrace,\lbrace e_1,e_5,e_9\rbrace,\lbrace e_4,e_8,e_9\rbrace,\lbrace e_2,e_6,e_{10}\rbrace,\lbrace e_3,e_7,e_{10}\rbrace\bigg\rbrace\ .
\end{aligned}
\end{equation}
So we have,
\begin{equation}
\begin{aligned}
&\mathcal{U}=\alpha_{10}\alpha_2\alpha_6 + \alpha_2\alpha_4\alpha_6 + \alpha_{10}\alpha_3\alpha_7 + \alpha_2\alpha_4\alpha_7 + \alpha_2\alpha_5\alpha_7 + \alpha_3\alpha_5\alpha_7\\
&\quad + \alpha_2\alpha_4\alpha_8 + \alpha_2\alpha_5\alpha_8 + \alpha_3\alpha_5\alpha_8 + \alpha_2\alpha_6\alpha_8 + \alpha_3\alpha_6\alpha_8 + \alpha_4\alpha_6\alpha_8\\
&\quad + \alpha_4\alpha_8\alpha_9 + \alpha_1 (\alpha_4 (\alpha_6 + \alpha_7) + \alpha_3 (\alpha_5 + \alpha_6 + \alpha_7) + \alpha_5 (\alpha_7 + \alpha_9))
\end{aligned}
\end{equation}
One can easily check that for $e=e_{10}$ we have,
\begin{equation}
\begin{aligned}
f_{e_{10},14}&=-\frac{\alpha_1\alpha_3 (\alpha_5+\alpha_6)+\alpha_2\alpha_6\alpha_8+\alpha_3\alpha_6\alpha_8+\alpha_2\alpha_4 (\alpha_7+\alpha_8)+\alpha_2\alpha_5 (\alpha_7+\alpha_8)+\alpha_3\alpha_5 (\alpha_7+\alpha_8)}{\mathcal{U}},\\
f_{e_{10},23}&=-\frac{\alpha_2\alpha_5\alpha_7+\alpha_2\alpha_4 (\alpha_6+\alpha_7)+\alpha_1 (\alpha_5\alpha_7+\alpha_3 (\alpha_6+\alpha_7)+\alpha_4 (\alpha_6+\alpha_7))+(\alpha_3+\alpha_4)\alpha_6\alpha_8}{\mathcal{U}},
\end{aligned}
\end{equation}
which implies that on the $+\alpha$-Landau surface $f_{e_{10},14}\neq0$ and $f_{e_{10},23}\neq0$.

\subsection{Diagram 11.b)}
For this diagram we get,
\begin{equation}
\begin{aligned}
&e_1=\ell_1,\ e_2=p_1+\ell_1,\ e_3=p_1+p_4+\ell_1,\ e_4=p_1+p_4+\ell_1+\ell_2,\ e_5=p_1+p_4+\ell_1+\ell_2-\ell_3,\\
&e_6=p_1+p_4+p_3+\ell_1+\ell_2-\ell_3,\ e_7=\ell_1+\ell_2-\ell_3,\ e_8=\ell_1-\ell_3,\ e_9=\ell_2,\ e_{10}=\ell_3.
\end{aligned}
\end{equation}
The non zero $\eta_{ve}$ and $\eta_{Ie}$ components are given by,
\begin{equation}
\begin{aligned}
&\eta_{v_1,e_1}=1,\ \eta_{v_1,e_2}=-1,\ \eta_{v_2,e_6}=1,\ \eta_{v_2,e_7}=-1,\ \eta_{v_3,e_5}=1,\ \eta_{v_3,e_6}=-1,\ \eta_{v_4,e_2}=1,\\ 
&\eta_{v_4,e_3}=-1,\ \eta_{v_5,e_1}=-1,\ \eta_{v_5,e_8}=1,\ \eta_{v_5,e_{10}}=1,\ \eta_{v_6,e_7}=1,\ \eta_{v_6,e_8}=-1,\ \eta_{v_6,e_{9}}=-1,\\
&\eta_{v_7,e_4}=1,\ \eta_{v_7,e_5}=-1,\ \eta_{v_7,e_{10}}=-1,\ \eta_{v_8,e_3}=1,\ \eta_{v_8,e_4}=-1,\ \eta_{v_8,e_9}=1\ .\\
\\
&\eta_{\ell_1,e_i}=1,\ \forall i=1,2,3,4,5,6,7,8\ ,\ \eta_{\ell_2,e_i}=1,\ \forall i=4,5,6,7,9\ ,\\
&\eta_{\ell_3,e_i}=1,\ \forall i=10\ \text{and}\ \eta_{\ell_3,e_i}=-1,\ \forall i=5,6,7,8\ .
\end{aligned}
\end{equation}
The spanning trees are,
\begin{equation}
\begin{aligned}
&\mathcal{T}^C=\bigg\lbrace\lbrace e_1,e_4,e_6\rbrace,\lbrace e_1,e_4,e_7\rbrace,\lbrace e_1,e_6,e_9\rbrace,\lbrace e_2,e_4,e_6\rbrace,\lbrace e_2,e_4,e_7\rbrace,\lbrace e_2,e_6,e_8\rbrace,\lbrace e_2,e_4,e_8\rbrace,\\
&\quad\lbrace e_2,e_5,e_8\rbrace,\lbrace e_2,e_5,e_9\rbrace,\lbrace e_2,e_6,e_9\rbrace,\lbrace e_2,e_6,e_{10}\rbrace,\lbrace e_2,e_7,e_{10}\rbrace,\lbrace e_2,e_9,e_{10}\rbrace,\lbrace e_3,e_5,e_{8}\rbrace,\\
&\quad\lbrace e_3,e_6,e_8\rbrace,\lbrace e_3,e_6,e_{10}\rbrace,\lbrace e_4,e_6,e_8\rbrace,\lbrace e_6,e_9,e_{10}\rbrace,\lbrace e_1,e_5,e_{9}\rbrace,\lbrace e_3,e_7,e_{10}\rbrace\bigg\rbrace\ .
\end{aligned}
\end{equation}
So we have,
\begin{equation}
\begin{aligned}
&\mathcal{U}=\alpha_2\alpha_4\alpha_6 + \alpha_2\alpha_4\alpha_7 + \alpha_1\alpha_4 (\alpha_6 + \alpha_7) + \alpha_2\alpha_4\alpha_8 + \alpha_2\alpha_5\alpha_8 \\
&\quad + \alpha_3\alpha_5\alpha_8 + \alpha_2\alpha_6\alpha_8 + \alpha_3\alpha_6\alpha_8 + \alpha_4\alpha_6\alpha_8 + \alpha_2\alpha_5\alpha_9 + \alpha_2\alpha_6\alpha_9 \\
&\quad + \alpha_1 (\alpha_5 + \alpha_6)\alpha_9 + \alpha_10 (\alpha_3 (\alpha_6 + \alpha_7) + \alpha_6\alpha_9 + \alpha_2 (\alpha_6 + \alpha_7 + \alpha_9))
\end{aligned}
\end{equation}
One can easily check that for $e=e_{10}$ in this case as well we have,
\begin{equation}
\begin{aligned}
f_{e_{10},14}&=-\frac{\alpha_2 (\alpha_4\alpha_8+(\alpha_5+\alpha_6) (\alpha_8+\alpha_9))}{\mathcal{U}},\\
f_{e_{10},23}&=-\frac{\alpha_6 (\alpha_4\alpha_8+(\alpha_1+\alpha_2) (\alpha_4+\alpha_9))}{\mathcal{U}},
\end{aligned}
\end{equation}
which implies that on the $+\alpha$-Landau surface $f_{e_{10},14}\neq0$ and $f_{e_{10},23}\neq0$.

\end{document}